\documentclass[aps,reprint,prx,twocolumn,superscriptaddress]{revtex4-2}

\usepackage{hyperref}
\usepackage{balance}
\usepackage{amssymb}
\usepackage{suffix}
\usepackage{mathtools}
\usepackage[utf8]{inputenc}
\usepackage{booktabs}
\usepackage{cases}
\usepackage[multiple]{footmisc}
\usepackage{dcolumn}
\usepackage{color,soul}
\usepackage{rotating}
\usepackage{perpage}
\usepackage{siunitx}
\usepackage{xcolor}
\usepackage{soul}
\usepackage{amsmath}
\usepackage{tikz}
\usepackage[T1]{fontenc}
\usepackage{etoolbox}
\usepackage{graphics}
\usepackage{siunitx}
\usepackage{float}	
\usepackage{collref}
\usepackage{multirow}
\usepackage{mathtools}
\usepackage{bm}
\usepackage{url}

\usepackage{tikz}
\usepackage{tikz-3dplot}
\usepackage{accents}

\DeclarePairedDelimiterX\MeijerM[3]{\lparen}{\rparen}%
{\begin{smallmatrix}#1 \\ #2\end{smallmatrix}\delimsize\vert\,#3}

\newcommand\MeijerG[8][]{%
	\mathbb{G}^{\,#2,#3}_{#4,#5}\MeijerM[#1]{#6}{#7}{#8}}

\WithSuffix\newcommand\MeijerG*[7]{%
	\mathbb{G}^{\,#1,#2}_{#3,#4}\MeijerM*{#5}{#6}{#7}}

\begin{document} 
	\title{Noncollinear twisted RKKY interaction on the optically driven SnTe(001) surface}
	\author{Mohsen Yarmohammadi}
	\email{mohsen.yarmohammadi@utdallas.edu}
	\address{Department of Physics, The University of Texas at Dallas, Richardson, Texas 75080, USA}
	\author{Marin Bukov}
	\email[]{mgbukov@phys.uni-sofia.bg}
	\affiliation{Max Planck Institute for the Physics of Complex Systems, N\"othnitzer Str.~38, 01187 Dresden, Germany}
	\affiliation{Faculty of Physics, St.~Kliment Ohridski University of Sofia, 5 James Bourchier Blvd, 1164 Sofia, Bulgaria}
	\author{Michael H. Kolodrubetz}
	\email{mkolodru@utdallas.edu}
	\address{Department of Physics, The University of Texas at Dallas, Richardson, Texas 75080, USA}
	\date{\today}
	
	\begin{abstract}
		The nontrivial spin texture on the (001) surface of topological crystalline insulator SnTe hosts exotic scientific importance and spintronic applications. Here, we study the effects of weak Floquet optical driving on the Ruderman-Kittel-Kasuya-Yosida (RKKY) interaction between two magnetic impurities on a doped SnTe(001) surface. Due to peculiar spin-orbit hybridization, we find a noncollinear twisted RKKY interaction comprising XYZ-Heisenberg, symmetric in-plane, and asymmetric Dzyaloshinskii-Moriya (DM) terms. We see that contributions from the $z$ ($x$)-component of the XYZ-Heisenberg (DM) interaction are dominant for most parameters. The interactions, including DM terms that are responsible for interesting spin textures, require doping in most cases. We propose to modify the interactions in situ via optical control of band structure, and thereby doping. A notable aspect of this control protocol is breaking of electron-hole symmetry, which stems from the DM interaction.
	\end{abstract}
	
	\maketitle
	{\allowdisplaybreaks
		
		
		\section{Introduction}
		
		For the past decade, topological insulators~(TIs) have been at the forefront of condensed matter physics~\cite{RevModPhys.82.3045,RevModPhys.83.1057}. Due to the strong spin-orbit coupling~(SOC)~\cite{Zhang2009,PhysRevB.75.121306}, topological phases of matter contain gapped bulk and gapless surface states, protected by the time-reversal symmetry. Notably, quantum anomalous Hall effect~(QAHE) has been observed in magnetically doped TIs~\cite{doi:10.1126/science.1187485,doi:10.1126/science.1234414,PhysRevLett.113.137201,Checkelsky2014}, which is a topological effect with potential for low-power information processing associated with dissipationless chiral edge transport~\cite{doi:10.1063/5.0100989}. However, the QAHE becomes weak in a wide variety of transition metal-doped TIs~\cite{PhysRevB.74.224418,PhysRevB.86.205127,PhysRevB.81.195203,Chang2015,PhysRevLett.113.137201} if one increases the density of surface carriers, mainly due to weakening of the magnetic order as a prerequisite for anomalous Hall physics. 
		
		Although the mechanism of this phenomenon is still under discussion~\cite{PhysRevB.94.195140}, it was suggested that the indirect Ruderman-Kittel-Kasuya-Yosida~(RKKY) exchange coupling between magnetic ions/impurities plays a significant role~\cite{PhysRevLett.115.036805,PhysRevLett.102.156603} among other common types of magnetism because it is tightly connected to the density of itinerant electrons in the host material~\cite{10.1143/PTP.16.45,*PhysRev.106.893,*PhysRev.96.99}. Depending on the spin-orbit structures of topological materials, various types of RKKY mechanisms~\cite{PhysRevB.87.125401,PhysRevB.87.045422,Dmitrienko2010} and eventually different weak/strong regimes of QAHE can emerge. Furthermore, emergent spintronic applications need the RKKY mechanism to find the proper material with improved adjustability of spin alignments for logic magnetic devices~\cite{Awschalom2007,RevModPhys.79.1217}. It is the latter purpose that we mainly aim at in this work, and we only conceptually~(i.e., without additional calculations) state its relationship with the QAHE.
		
		It has been shown that nontrivial RKKY interactions can appear on the surface of three-dimensional TIs~\cite{PhysRevLett.102.156603,PhysRevB.81.172408,PhysRevLett.106.136802,PhysRevB.89.115431} as well as at the edge of two-dimensional TIs~\cite{PhysRevB.80.241302,PhysRevB.94.155408,PhysRevLett.120.147201}. In this work, we proceed with the topological crystalline insulators~(TCIs)~\cite{PhysRevLett.106.106802,Hsieh2012,Tanaka2012,Dziawa2012,Xu2012} because, in addition to the time-reversal symmetry, crystal symmetries protect gapless surface states as well. This, in turn, covers a broad range of materials rather than specific TI compounds. Moreover, in contrast to an odd number of Dirac cones in strong TIs, TCIs provide an even number of cones, useful for spintronics and valleytronics~\cite{Awschalom2007,RevModPhys.79.1217,PhysRevLett.116.016802}. We choose to focus on SnTe(001) surface as one of the well-known TCIs with a \textit{nontrivial spin texture}~\cite{PhysRevB.87.235317} to see how its magnetic features can go beyond the ones in TIs to be reflected in spintronics and QAHE physics.
		
		In order to understand novel features in these materials, we need the ability to tune the properties of the surface. The best way to do so is through a gap opening at Dirac cones because it suppresses the large density of the doped surface around the Fermi energy, first, and second, it makes the magnetic order strong enough through the suppression of various phase alterations of ferromagnetic~(FM) and antiferromagnetic~(AFM) interactions. Among various ways for gap opening~\cite{PhysRevB.89.195413}, we proceed with optical driving using Floquet-Bloch states~\cite{PhysRevB.79.081406,doi:10.1146/annurev-conmatphys-031218-013423,PhysRevB.84.235108,PLATERO20041} because by tuning the intensity and frequency of the radiation, one can dynamically tune the photon-dressed bands, and eventually modulate the properties of an irradiated system in a controlled way. 
		
		Motivated by various light-induced phenomena~\cite{nasu2004photoinduced,PhysRevLett.117.087402,PhysRevB.94.041409,PhysRevB.96.041205,PhysRevLett.111.136402,PhysRevLett.126.086801}, we consider the interaction of a weak~(to avoid the heating issue) circularly polarized light with the SnTe(001) surface, which opens a gap isotropically at all four Dirac cones, and thereby modulates the RKKY interaction. Subsequently, assuming that particles fill the Floquet-Bloch bands up to a given chemical potential when the weak Floquet drive is turned on, we pose the question: how does the interplay between doping and optical driving influence the RKKY interaction for both spintronics and QAHE treatment? To answer this question, we reformulate the Heisenberg RKKY interaction~\cite{PhysRevB.102.075411,SR} of hybrid states of sublattices in SnTe(001) for the bare spin-orbital states of each sublattice to take into account the effect of a nontrivial spin texture. This results in \textit{noncollinear twisted spin alignments} including XYZ-Heisenberg, symmetric in-plane, and asymmetric DM interactions, implying that magnetic moments lie on the surface with increased adjustability. The asymmetric DM interaction is substantial for spintronic applications~\cite{Fert2013,PhysRevLett.119.176809,Wakatsuki2015}. Interestingly, the Floquet drive tends to treat the QAHE by suppressing the density of doped states and RKKY amplitudes; however, we note that topological features of the SnTe(001) surface are not essential for our study. 
		
		This paper is structured as follows. In Sec.~\ref{s2}, we present the continuum Hamiltonian of pristine SnTe(001) surface and construct the main building block of optically driven states. In Sec.~\ref{s3}, we turn to the RKKY theory and derive the Floquet-Green's functions for a weak and off-resonant drive. The numerical results are presented in Sec.~\ref{s4}. We discuss potential extensions in Sec.~\ref{s5n} and finally we end the paper with a conclusion in Sec.~\ref{s6}.
		
		
		\section{Hamiltonian model}\label{s2}
		
		\subsection{Pristine $\rm SnTe$(001)}\label{s2_1}
		In this section, we intorduce the continuum Hamiltonian model~\cite{Tanaka2012,Xu2012,Dziawa2012,Hsieh2012,PhysRevB.88.241303,PhysRevB.89.195413} to describe each coaxial Dirac cone in the low-energy limit of SnTe(001) surface. Two cones $\Lambda_x$ and $\Lambda'_x$ are located along the direction $X_1-\Gamma-X_1$ of the projected surface Brillouin zone~(SBZ), while two cones $\Lambda_y$ and $\Lambda'_y$ are located along the direction $X_2-\Gamma-X_2$, as shown in the left side of Fig.~\ref{fig1}. The pristine Hamiltonian of the $\Lambda_x$ point is given by~($\hbar = 1$)~\cite{PhysRevB.89.195413,PhysRevB.90.035402,PhysRevB.102.075411,PhysRevB.88.241303,SR,Okada1496}
		\begin{equation}\label{eq_1}
			\mathcal{H}_{\Lambda_x} = \tilde{\eta}_x  p_x \sigma_y - \tilde{\eta}_y p_y \sigma_x \, ,
		\end{equation}
		where $p_x = k_x - \Lambda_x$ is measured from the Dirac cone at $\Lambda_x = ( \sqrt{n^2+\delta^2}/\eta_x,0)$, $p_y = k_y$, $\tilde{\eta}_x = (\delta/\sqrt{n^2+\delta^2})\eta_x$, and $\tilde{\eta}_y = \eta_y$~\cite{PhysRevB.90.035402}. Here, $\eta_x = 3.53$ eV.\AA\, and $\eta_y = 1.91$ eV.\AA\, refer to the bare Fermi velocities along the $x$- and $y$-direction, respectively. The Pauli matrices are given by $\sigma_x$ and $\sigma_y$. The intervalley scattering parameters $n = 55$ meV and $\delta = 40$ meV are, respectively, responsible for the high-energy Dirac cones and the formation of two copies of Dirac cones with opposite chiralities in the four-band model~(not shown here)~\cite{Hsieh2012,PhysRevB.89.195413,PhysRevB.88.241303,PhysRevB.102.075411}. Although they correctly describe intervalley scattering at the lattice scale of SnTe(001), they vary sample to sample since they arise from non-universal effects such as surface roughness. Indeed, the parameter $n$ is the threshold energy below which the low-energy model is valid. 
		\begin{figure}[t]
			\centering
			\includegraphics[width=1\linewidth]{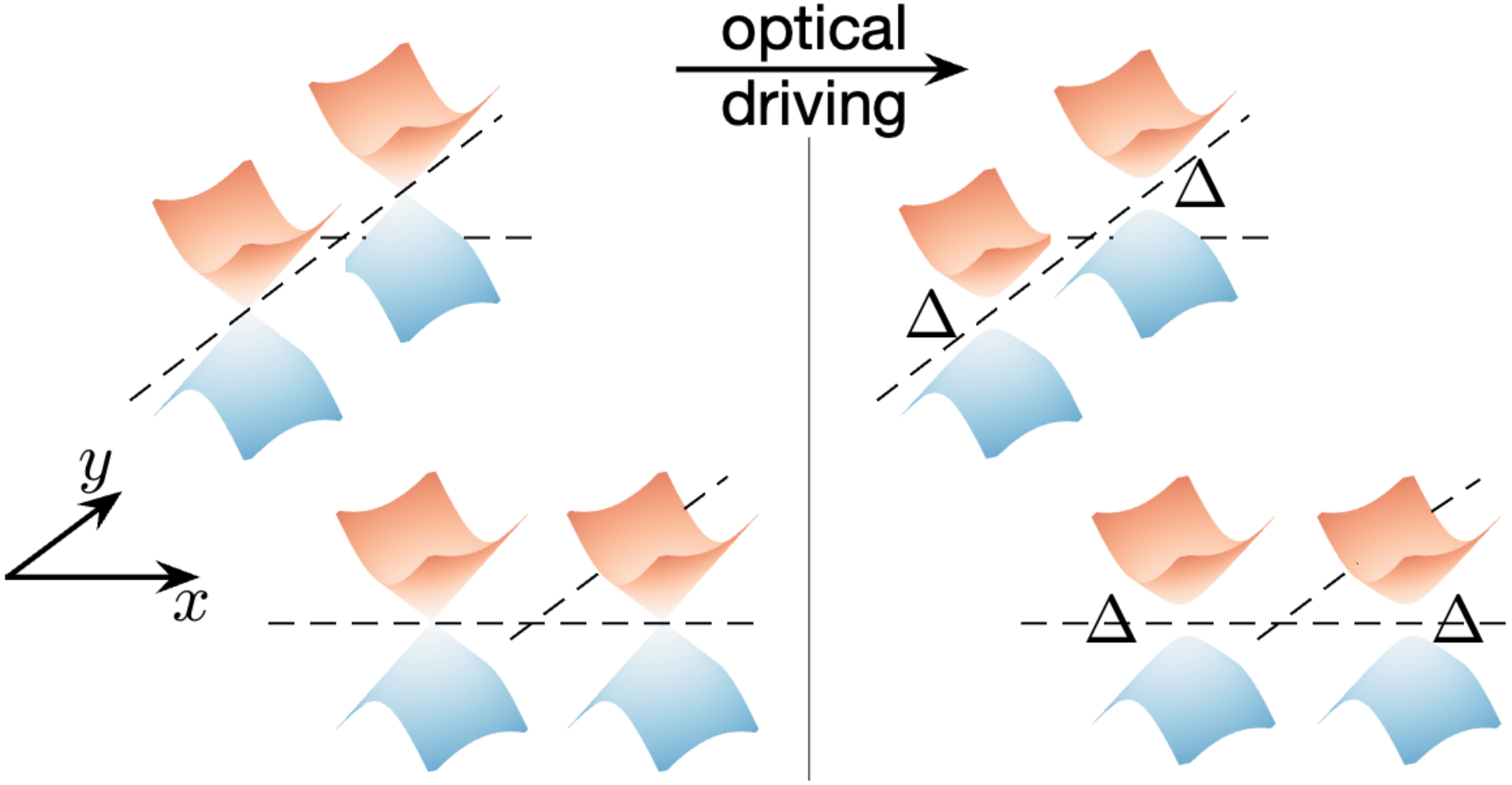}
			\caption{The low-energy spectrum of Dirac fermions on the SBZ of SnTe(001) surface, left side, and the optically driven spectrum, right side, with induced gap $\Delta = 2 a^2/\Omega$ from the light with the ac field $a$ and the frequency $\Omega$. Other Dirac cones follow from $C_2$ and $C_4$ rotational symmetries.} 
			\label{fig1}
		\end{figure}
		
		The rotational symmetry $C_2$ protects the coaxial Dirac cones resulting in $\mathcal{H}_{{\Lambda'_{x,y}}} = \mathcal{H}_{{\Lambda_{x,y}}}$, while the perpendicular cones are formed by the $C_4$ symmetry. This, in turn, implies that $	\mathcal{H}_{\Lambda_y} = {} \tilde{\eta}_y  p_x \sigma_y - \tilde{\eta}_x p_y \sigma_x$. It should be stressed that the valence and conduction bands of the energy spectrum refer to the hybridized $p$-orbitals of the cation Sn$^{2+}$ and anion Te$^{2-}$ sublattices, namely $|1\rangle = (|\uparrow, \rm Te\rangle+|\downarrow, \rm Sn\rangle)/\sqrt{2}$ and $|2\rangle = (|\uparrow, \rm Sn\rangle+|\downarrow, \rm Te\rangle)/\sqrt{2}$~\cite{PhysRevB.87.235317,PhysRevB.89.195413,PhysRevB.90.035402,PhysRevB.88.241303}. 
		
		Before turning to the irradiated SnTe(001) surface, we make one useful simplification, which will help later to find analytical RKKY interactions. Because RKKY theory is mainly based on low-momenta response, we neglect the anisotropy feature of the surface Dirac cones originating from $\tilde{\eta}_x \simeq 2.08$ eV.\AA\, and $\tilde{\eta}_y = 1.91$ eV.\AA. Hence, we set $v_{\rm F} = (\tilde{\eta}_x + \tilde{\eta}_y )/2 \simeq 2$ eV.\AA.
		
		
		\subsection{Irradiated $\rm SnTe$(001)}\label{s2_2}
		
		In this section, we present the expression for the effective two-band Hamiltonian model discussed above for the case when the SnTe(001) Dirac cones are \textit{weakly} driven by a circularly polarized light. To do so, the time-periodic vector potential $\vec{A}(t) =  A_0 [\sin(\Omega t),\cos(\Omega t)]$, where $A_0 = E_0/\sqrt{2} \Omega$ ($E_0$ and $\Omega$ are the amplitude and frequency of light, respectively) is chosen to drive the Hamiltonian of the $\Lambda_x$ point (and other Dirac cones) through the minimal coupling scheme as
		\begin{equation}\label{eq_2}
			\mathcal{H}_{\Lambda_x}(t) = {} 
			\begin{pmatrix}
				0 & - {\tt i} v_{\rm F}p_- - a e^{+{\tt i} \Omega t}\\\\
				+{\tt i} v_{\rm F}p_+ - a e^{-{\tt i} \Omega t} & 0
			\end{pmatrix} ,
		\end{equation}
		where $p_\pm = p_x \pm {\tt i} p_y$ and $a = e A_0 v_{\rm F}$. The drive induces transitions between the eigenstates. Since the Hamiltonian is periodic in time, $\mathcal{H}_{\Lambda_x}(t) = \mathcal{H}_{\Lambda_x}(t + T)$ with $T = 2 \pi/\Omega$, the valid theory to address the time-periodic Hamiltonians is the Floquet-Bloch's theorem~\cite{GRIFONI1998229,PLATERO20041}: a complete set of solutions of the time-dependent Schr\"odinger equation $\mathcal{H}_{\Lambda_x}(t) |\psi(t) \rangle = {\tt i} \partial_t |\psi(t) \rangle$ reads as $	|\psi_{f,p}(t) \rangle = {} e^{-{\tt i} \varepsilon_{f,p}} |\mathcal{F}_{f,p}(t) \rangle$, where the integer Floquet index $f$ classifies different sidebands with energies $\varepsilon_{f,p}$. Accordingly, the Fourier series $	|\mathcal{F}_{f,p}(t) \rangle = \sum^\infty_{j=-\infty} e^{-{\tt i}j\Omega t} | \mathcal{F}^j_{f,p}\rangle$ describes the time-periodic Floquet states $|\mathcal{F}_{f,p}(t) \rangle = |\mathcal{F}_{f,p}(t + 2 \pi/\Omega) \rangle$. It should be mentioned that $f = 0$ maps the quasienergies to the first Floquet zone $[-\Omega/2,\Omega/2]$. Then, it is straightforward to deduce the Floquet Hamiltonian $\mathcal{H}_{{\rm F},\Lambda_x} = \mathcal{H}_{\Lambda_x}(t) - {\tt i} \partial_t$ in the Fourier domain~\cite{PhysRev.138.B979} $\sum^\infty_{j=-\infty} \mathcal{H}_{{\rm F},m j} | \mathcal{F}^j_{f,p}\rangle = \varepsilon_{f,p} | \mathcal{F}^m_{f,p}\rangle$. Having the driving period $T = 2 \pi/\Omega$, the matrix elements of the Floquet Hamiltonian can be found via $\mathcal{H}_{{\rm F},m j} = {} \mathcal{H}_{m j} - j \Omega \delta_{m,j}$, where $j$ is the photon number and~\cite{PhysRevB.85.155449,PhysRevB.78.235124,PhysRevResearch.2.033228} 
		\begin{equation}\label{eq_4}
			\mathcal{H}_{m j} = {} \frac{1}{T} \int^T_0 \mathcal{H}_{\Lambda_x}(t) e^{{\tt i} (m-j) \Omega t} dt\, .
		\end{equation}
		
		The Floquet Hamiltonian can be numerically diagonalized to study the exact dynamics governed by the application of light. However, for sufficiently weak electric field intensities, i.e., when the field intensity $a$ is small compared to the frequency $\Omega$, we may approximate the Floquet Hamiltonian using a high-frequency expansion explained below. In the high-frequency limit, the Floquet sidebands do not overlap; thus, there are no gaps due to band crossings so the hybridization between different Floquet bands becomes weak, meaning that interband transitions of electrons are suppressed.
		
		Thus, using a perturbative approach known as the van Vleck inverse-frequency expansion, we obtain the effective Floquet Hamiltonian in powers of $\Omega^{-1}$, $\mathcal{H}_{{\rm F},\Lambda_x} \simeq {} \mathcal{H}_{\Lambda_x} + \frac{\left[\mathcal{H}_{-1},\mathcal{H}_{+1}\right]}{\Omega}$~\cite{PhysRevB.84.235108,PhysRevB.82.235114,PhysRevB.84.235108,PhysRevLett.110.026603}, which reads as:
		\begin{equation}\label{eq_5}
			\begin{aligned}
				\mathcal{H}_{{\rm F},\Lambda_x} = {} & \begin{pmatrix}
					+ \Delta/2& - {\tt i} v_{\rm F}\,p\, e^{-{\tt i} \varphi_p}\\\\
					+ {\tt i} v_{\rm F}\,p\, e^{+{\tt i} \varphi_p} & -\Delta/2
				\end{pmatrix}\, ,
			\end{aligned}
		\end{equation}
		where a gap $\Delta = {} 2 a^2/\Omega$ opens up at the Dirac cone through a two-photon process similar to that in Refs.~\cite{PhysRevB.79.081406,doi:10.1063/1.3597412,PhysRevB.83.245436,PhysRevB.85.155449,PhysRevResearch.2.033228}. Using the $C_2$ and $C_4$ symmetries, one can obtain the Floquet Hamiltonian at other Dirac cones. Diagonalizing the above Hamiltonian leads to a gapped Dirac dispersion $\mathcal{E}_{\Lambda_x}(p) = {} \pm\, \sqrt{v^2_{\rm F} p^2 + \Delta^2/4}$, as shown in the right panel of Fig.~\ref{fig1}. 
		
		In addition to optical driving, there are various ways to open the gap at the Dirac cones~\cite{PhysRevB.89.195413}, namely the exchange magnetization, the Zeeman term without an external magnetic field, and proximity coupling to a ferromagnet. However, as discussed in the introduction, the Floquet drive allows us to tune the properties of the dispersion for various purposes by adjusting the intensity and frequency of light. Moreover, further photon processes in $\mathcal{H}_{{\rm F},\Lambda_x}$ can be taken into account by including higher-order terms~\cite{PhysRevB.85.155449,PhysRevB.85.155449,PhysRevResearch.2.033228}. 
		
		For our purposes, the Floquet drive will be used to tune the RKKY interaction for both spintronics and eventually QAHE physics (indirectly through the density of surface carriers). We now proceed to the RKKY interaction in irradiated SnTe(001).
		
		
		\section{RKKY interaction in irradiated $\rm SnTe$(001)}\label{s3}
		As explained in the introduction, the indirect exchange coupling between two magnetic impurities or two localized spins $\vec{S}_1$ and $\vec{S}_2$ mediated by the host itinerant electron spins $\vec{s}$ is described by RKKY theory~\cite{10.1143/PTP.16.45,*PhysRev.106.893,*PhysRev.96.99}. In our case, the host itinerant electrons are the optically driven ones on the SnTe(001) surface given by Eq.~\eqref{eq_5}. According to this theory, the magnetic moments are treated as immobile defects on the lattice sites $\vec{R}_1$ and $\vec{R}_2$, and the interaction Hamiltonian is given by $\mathcal{H}_{\rm int} = J \sum_i \vec{S}_i \cdot \vec{s}_i$, where $J$ denotes the bare exchange energy. Thus, we consider the total Hamiltonian for SnTe(001) surface including a contact interaction with the magnetic centers, namely, $\mathcal{H} = \mathcal{H}_{{\rm F},\Lambda_x} + \mathcal{H}_{\rm int}$. Using second-order perturbation theory, one finds~\cite{10.1143/PTP.16.45,*PhysRev.106.893,*PhysRev.96.99}
		\begin{equation}\label{eq_6}
			\mathcal{H}^{\alpha \beta}_{\rm RKKY} = {} J^2 \sum_{l,j}\, S^{l\alpha}_1 \,\chi^{\alpha \beta}_{lj}(\,\vec{R}\,) \, S^{j\beta}_2\, , 
		\end{equation}with lattice sites $\alpha$ and $\beta$, and the impurity separation $\vec{R} = \vec{R}_2 - \vec{R}_1$. In the above equation, the spin susceptibility $\chi^{\alpha \beta}_{lj}(\,\vec{R}\,)$ can be obtained from the retarded Green's functions in the spin-space for different flavors $\{l,j\} \in \{x,y,z\}$~\cite{Kogan,PhysRevB.83.165425,PhysRevB.84.125416}: 
		\begin{equation}\label{eq_7}
			\chi^{\alpha \beta}_{lj}(\,\vec{R}\,) = -\frac{2}{\pi}\, \Im \int^{\mathcal{E}_{\mathrm{F}}}_{-\infty} d\mathcal{E}\, \mathbb{F}^{\alpha \beta}_{lj}(\mathcal{E},\vec{R})\, ,
		\end{equation}
		where $\mathcal{E}_{\rm F}$ is the Fermi energy and
		\begin{equation}\label{eq_8}
			\mathbb{F}^{\alpha \beta}_{lj}(\mathcal{E},\vec{R})=\mathrm{Tr}\left[\sigma_{l}  G^{\alpha \beta}(\mathcal{E},\vec{R}) \sigma_{j}  G^{\beta \alpha}(\mathcal{E},-\vec{R}) \right]\, .
		\end{equation}
		
		We set the temperature to zero in the present work, so $\mathcal{E}_{\rm F} = \mu$, where $\mu$ is the chemical potential. The role of doping is then characterized by $\mu>0$ and $\mu <0$ referring to electron and hole doping, respectively. We assume that the electrons fill all energies up to $\mu$ within the Floquet-modified dispersion because the optical drive considered is sufficiently weak and off-resonant that it may be turned on continuously, resulting in a new (quasi zero temperature) steady state. Note that this is fundamentally different from many other Floquet systems which are experimentally constrained to pump-probe non-equilibrium states to avoid heating, for which the electron filling is quite different~\cite{Rudner2020}. In Eq.~\eqref{eq_8}, we have
		\begin{equation}\label{eq_9}
			G^{\alpha \beta}(\mathcal{E},\vec{R}) = \begin{pmatrix}
				G^{\uparrow \uparrow}_{\alpha \beta}\,(\mathcal{E},\vec{R}) && G^{\uparrow \downarrow}_{\alpha \beta}\,(\mathcal{E},\vec{R})\\\\
				G^{\downarrow \uparrow}_{\alpha \beta}\,(\mathcal{E},\vec{R}) && G^{\downarrow \downarrow}_{\alpha \beta}\,(\mathcal{E},\vec{R})
			\end{pmatrix}\, ,
		\end{equation}
		which denotes the real-space non-interacting Green’s function of sublattices in spin-space, given by the photo-dressed Dirac bands. However, we must sum over the directions $X_1-\Gamma-X_1$ and $X_2-\Gamma-X_2$ on the SnTe(001) surface, as well as the pair of Dirac cones at $X_1$ and $X_2$ points, to cover the entire SBZ. Thus, we have
		\begin{equation}\label{eq_10}
			\begin{aligned}
				G^{\sigma \sigma'}_{\alpha \beta}(\mathcal{E},\vec{R}) =&\frac{1}{S_{\mathrm{SBZ}}}\int d^2p\, e^{\texttt{i}\vec{p}\cdot \vec{R}} \Big[e^{\texttt{i}\vec{X}_1\cdot \vec{R}} G^{\sigma \sigma'}_{\alpha \beta}(\vec{p}+ \vec{X}_1,\mathcal{E})  \\ {} & + e^{\texttt{i}\vec{X}_2\cdot \vec{R}} G^{\sigma \sigma'}_{\alpha \beta}(\vec{p} + \vec{X}_2,\mathcal{E})\Big]\, ,
			\end{aligned}
		\end{equation}with $\{\sigma, \sigma'\} = \{\uparrow,\downarrow\}$, where $S_{\mathrm{SBZ}}$ is the area of the entire SBZ. The integral is dominated by low momentum, $p \ll |\Lambda_{x,y}|$, justifying our choice to treat each Dirac cone independently in the low-energy limit of two-band model.
		
		It is necessary to point out that in the Hamiltonian model, a superposition of sublattices given by the eigenstates $|1\rangle$ and $|2\rangle$ appears, rather than the bare sublattices $|{\rm Te/Se}, \uparrow/\downarrow\rangle $. Therefore, one needs to express the real-space Green's functions of the hybrid states in terms of the bare sublattices considering the corresponding spins. Hence, we rewrite the above Green's function as~[see Appendix~\ref{ap_1}]
		\begin{equation}\label{eq_11}
			G^{\sigma \sigma'}_{\alpha \beta}(\mathcal{E},\vec{R}) = \sum_{\gamma,\eta} u^*_{\alpha,\gamma,\sigma}u_{\beta,\gamma,\sigma'} \mathcal{G}_{\gamma \eta}(\mathcal{E},\vec{R})\, ,
		\end{equation}
		in which the summation runs over hybrid states $\{|1\rangle,|2\rangle\}$ and $u_{\rm Te,1,\uparrow} = u_{\rm Te,2,\downarrow} = u_{\rm Sn,1,\downarrow} = u_{\rm Sn,2,\uparrow} = 1/\sqrt{2}$. To calculate $\mathcal{G}_{\gamma \eta}(\mathcal{E},\vec{R}) $, we first calculate the reciprocal-space Green's functions around all Dirac cones within the weak-driven approximation. Therefore, using the relation $\mathcal{G}^{\Lambda_x}(p,\mathcal{E}) = {}\left[\mathcal{E}+{\tt i}o^+ -\mathcal{H}_{{\rm F},\Lambda_x}\right]^{-1}$ with infinitesimal parameter $o^+ \to 0$ for the $\Lambda_x$ point as well as using the expressions $d^2p = p\,dp\,d\varphi_p$ and $\exp\big[\texttt{i}\vec{p}\cdot \vec{R}\big]=\exp\left[\texttt{i}\,p\,R\,\cos\left(\varphi_p-\varphi_R\right)\right]$ with $\varphi_p=\tan^{-1}\left(p_y/p_x\right)$, we obtain
		\begin{equation}\label{eq_12}
			\begin{aligned}
				\mathcal{G}_{\gamma \eta}(\mathcal{E}, \vec{R}) ={} \frac{1}{S_{\mathrm{SBZ}}} \int^{p_{\rm c}}_{0}  \int^{2\pi}_{0} d^2p
				{} e^{\texttt{i}\,\vec p\cdot \vec R}\, \mathcal{G}_{\gamma \eta}(p,\mathcal{E}) .
			\end{aligned}
		\end{equation}
		We set the cutoff $p_{\rm c} \to \infty$, as the long-range interactions are dominated by low momentum modes within this integral. Hence, for the $\Lambda_x$ point, the real-space Green's functions of hybrid states $|1\rangle$ and $|2\rangle$ read as
		\begin{subequations}\label{eq_13}
			\begin{align}
				\hspace*{-0.15cm}\mathcal{G}_{11}(\mathcal{E}, \vec{R}) = {} & \frac{-2\pi\left[\mathcal{E}+{\tt i}o^++\Delta/2\right]}{S_{\mathrm{SBZ}}\,v^2_{\rm F}} \mathbb{K}_0(-{\tt i}\widetilde{\mathcal{E}}\,R/v_{\rm F})\, ,\\
				\hspace*{-0.15cm}\mathcal{G}_{12}(\mathcal{E}, \vec{R}) = {} & \frac{+2\pi\,{\tt i}\,e^{-{\tt i}\varphi_R}\widetilde{\mathcal{E}}}{S_{\mathrm{SBZ}}\,v^2_{\rm F}} \mathbb{K}_1(-{\tt i}\widetilde{\mathcal{E}}\,R/v_{\rm F})\, ,\\
				\hspace*{-0.15cm}	\mathcal{G}_{21}(\mathcal{E}, \vec{R}) = {} & \frac{-2\pi\,{\tt i}\,e^{+{\tt i}\varphi_R}\widetilde{\mathcal{E}}}{S_{\mathrm{SBZ}}\,v^2_{\rm F}} \mathbb{K}_1(-{\tt i}\widetilde{\mathcal{E}}\,R/v_{\rm F})\, ,\\
				\hspace*{-0.15cm}	\mathcal{G}_{22}(\mathcal{E}, \vec{R}) = {} & \frac{-2\pi\left[\mathcal{E}+{\tt i}o^+-\Delta/2\right]}{S_{\mathrm{SBZ}}\,v^2_{\rm F}} \mathbb{K}_0(-{\tt i}\widetilde{\mathcal{E}}\,R/v_{\rm F})\, ,
			\end{align}
		\end{subequations}	where $\widetilde{\mathcal{E}} = \sqrt{(\mathcal{E}+{\tt i}o^+)^2 -\Delta^2/4}$ and $\mathbb{K}_{0,1}$ denote modified Bessel functions. For the $\Lambda'_x$, $\Lambda_y$, and $\Lambda'_y$ points, we have the exact same expressions since the gap is opened isotropically at all Dirac cones by the optical driving. The above elements fulfil the relations $\mathcal{G}_{11/22}(\mathcal{E},-\vec{R}) = \mathcal{G}_{11/22}(\mathcal{E}, \vec{R})$ and $\mathcal{G}_{12/21}(\mathcal{E}, -\vec{R}) = - \mathcal{G}_{12/21}(\mathcal{E}, \vec{R}) $.
		
		Turning back to Eq.~\eqref{eq_6}, one would rewrite the RKKY Hamiltonian as
		\begin{equation}\label{eq_14}
			\mathcal{H}_{\rm RKKY} = {}  - \frac{2J^2}{\pi} \text{Im} 	\int_{-\infty}^{\mu} d \mathcal{E}\sum_{lj}S^{l}_1 S^{j}_2 \, \mathbb{F}_{lj} \, .
		\end{equation}
		The components $\mathbb{F}^{\rm  \alpha\, \alpha}_{lj}$ and $\mathbb{F}^{\rm  \alpha\, \beta}_{lj}$ for the spins on the same and different sublattices, respectively, are presented in Appendix~\ref{ap_2}. Accordingly, the Hamiltonians read as \begin{subequations}\label{eq_15}
			\begin{align}
				\mathcal{H}^{ \alpha\, \alpha,\alpha_{\rm s}}_{\rm RKKY} = &  \sum_{i}\mathcal{J}^{ \alpha\, \alpha}_{i}S^i_1S^i_2 \hspace*{-0.05cm}+ \hspace*{-0.05cm} \vec{\mathcal{J}}_{\rm DM}^{\alpha_{\rm s}} \cdot (\vec{S}_1 \times \vec{S}_2) \notag \\  {} & + \alpha_{\rm s} \mathcal{J}^{ \alpha\, \alpha}_{xy} [S^x_1S^y_2 +S^y_1S^x_2]  \,,\label{eq_15a}\\
				\mathcal{H}^{ \alpha\, \beta,\alpha_{\rm d}}_{\rm RKKY} = & \sum_{i}\mathcal{J}^{ \alpha\, \beta}_{i}S^i_1S^i_2 \hspace*{-0.05cm} + \hspace*{-0.05cm}\vec{\mathcal{J}}_{\rm DM}^{\alpha_{\rm d}} \cdot (\vec{S}_1 \times \vec{S}_2) ,\label{eq_23nb}
			\end{align}
		\end{subequations}
		where we have $\vec{\mathcal{J}}^{\alpha_{\rm s}}_{\rm DM} =(\mathcal{J}^{ \alpha\, \alpha}_{{\rm DM},x},\alpha_{\rm s}\mathcal{J}^{ \alpha\, \alpha}_{{\rm DM},y},0)$, and $\vec{\mathcal{J}}^{\alpha_{\rm d}}_{\rm DM} =(\mathcal{J}^{ \alpha\, \beta}_{{\rm DM},x},0,\alpha_{\rm d}\mathcal{J}^{ \alpha\, \beta}_{{\rm DM},z})$. When the impurities are placed on the TeTe~(SnSn) sublattices, we use $\alpha_{\rm s} = + 1~(-1)$, while $\alpha_{\rm d} = + 1~(-1)$ for the TeSn~(SnTe) sublattices. 
		
		The first terms in both Hamiltonians of Eq.~\eqref{eq_15} couple the same spin directions with different exchange strengths \{$\mathcal{J}_x$, $\mathcal{J}_y$, $\mathcal{J}_z$\}, which gives the XYZ-Heisenberg interaction. Furthermore, due to the intrinsic SOC of the SnTe(001) surface, the symmetry of spin space is expected to be broken in response to the impurities, leading to off-diagonal components of the RKKY Hamiltonian~(the last terms)~\cite{PhysRevB.69.121303,PhysRevLett.106.097201}. While the asymmetric second terms resemble a DM interaction, the third term in Eq.~\eqref{eq_15a} gives rise to an in-plane symmetric interaction. The large DM interaction~(which we will show in the following) compared to the one in TI thin films~\cite{PhysRevB.96.024413} can lead to
		interesting phenomena such as spin Nernst effect, the appearance of the nontrivial topology, and topological spin excitations~\cite{PhysRevLett.117.217202,PhysRevLett.117.227201,PhysRevX.8.041028}. The DM interaction has been also explored extensively in recent years for spintronic applications~\cite{PhysRevLett.119.176809,Wakatsuki2015,Fert2013,Sampaio2013}. These off-diagonal terms enable the full RKKY Hamiltonian to provide noncollinear twisted alignment between the magnetic impurities. 
		
		Let us define the form factor between Dirac points and $\{X_1$  and $ X_2\}$ points from Eq.~\eqref{eq_10}
		\begin{equation}\label{eq_15n}
			f(\vec R) = 2\,[e^{{\tt i} X_1 R_x} \cos(\Lambda_x R_x) + e^{{\tt i} X_2 R_y} \cos(\Lambda_y R_y)] \, .
		\end{equation} 
		For clarity, we work with the normalized exchange interactions, $\mathcal{C}\,J^2\,|f(\vec R)|^2$, where $\mathcal{C} = 4\pi/S^2_{\rm SBZ} v^4_{\rm F}$ thus, $\widetilde{\mathcal{J}}(R) = \mathcal{J}(R)/\mathcal{C}\,J^2\,|f(\vec R)|^2$ suppresses the beating type of oscillations from the multiple surface Dirac cones~\cite{PhysRevB.102.075411}, as shown in Fig.~\ref{fc1} of Appendix~\ref{ap_1n}. With tthis, we show the results for a single Dirac cone in what follows. The modulations captured in $|f(\vec{R})|^2$ are different than those of graphene and other systems~\cite{PhysRevB.100.014410,PhysRevB.74.224438,PhysRevB.76.184430,PhysRevB.81.205416} and can be useful to modulate the couplings on medium-length scales $1/\Lambda_{x,y}$. Crucially, such a length scale only exists in TCIs and is potentially tunable via, e.g., strain. 
		
		For numerical purposes, we first define $\mathcal{E} - \mu = \mathcal{E}'$, owing to the electron-hole symmetry~\cite{PhysRevB.76.184430,PhysRevB.83.165425,PhysRevB.84.125416,Kogan}, to set the upper limit of the integral in Eq.~\eqref{eq_14} to zero. Then, we use the imaginary energy representation $\mathcal{E}' + {\tt i} o^+= {\tt i} \omega$ to get rid of the numerical singularities and tight tolerances in the real energy integration. After simple algebra calculations and defining $\widetilde{\omega}^2 = (\omega-{\tt i}\mu)^2+(\Delta^2/4)$, the components of RKKY Hamiltonians for the first configuration (impurities on the same sublattices) read as:
		\begin{subequations}\label{eq_16}
			\begin{align}
				\widetilde{\mathcal{J}}^{ \alpha\, \alpha}_{x}(R)= & +\text{Im} \int_{o^+}^{\infty} \hspace*{-0.1cm}{\tt i} d \omega \,\widetilde{\omega}^2 \big[\mathbb{K}^2_0(\widetilde{\omega}\,R/v_{\rm F}) \notag \\ {} & \qquad  +\cos(2 \varphi_R) \mathbb{K}^2_1(\widetilde{\omega}\,R/v_{\rm F})\big],\label{eq_16na}\\
				\widetilde{\mathcal{J}}^{ \alpha\, \alpha}_{y}(R) = & +\text{Im} \int_{o^+}^{\infty} \hspace*{-0.1cm}{\tt i} d \omega\, \widetilde{\omega}^2 \big[\mathbb{K}^2_0(\widetilde{\omega}\,R/v_{\rm F}) \notag \\ {} & \qquad  - \cos(2 \varphi_R) \mathbb{K}^2_1(\widetilde{\omega}\,R/v_{\rm F})\big],\\
				\widetilde{\mathcal{J}}^{ \alpha\, \alpha}_{z}(R) = & +\text{Im} \int_{o^+}^{\infty} \hspace*{-0.1cm}{\tt i} d \omega \,\widetilde{\omega}^2 \big[\mathbb{K}^2_0(\widetilde{\omega}\,R/v_{\rm F}) \notag \\ {} & \qquad  \hspace*{-2.5cm}+ \mathbb{K}^2_1(\widetilde{\omega}\,R/v_{\rm F})\big]-\frac{\Delta^2}{2} \text{Im} \int_{o^+}^{\infty} \hspace*{-0.1cm}{\tt i} d \omega \mathbb{K}^2_0(\widetilde{\omega}\,R/v_{\rm F}),\\
				\widetilde{\mathcal{J}}^{ \alpha\, \alpha}_{xy}(R) = & +\sin(2 \varphi_R) \text{Im} \int_{o^+}^{\infty} \hspace*{-0.1cm}{\tt i} d \omega\, \widetilde{\omega}^2\,\mathbb{K}^2_1(\widetilde{\omega}\,R/v_{\rm F}),\label{eq_16nd}\\
				\widetilde{\mathcal{J}}^{ \alpha\, \alpha}_{{\rm DM},x}(R) = & -2 \sin(\varphi_R)\,  \text{Im} \int_{o^+}^{\infty} \hspace*{-0.1cm} d \omega \,\widetilde{\omega} \sqrt{ \widetilde{\omega}^2-\Delta^2/4}\notag \\ {} & \hspace*{1cm} \,\times\mathbb{K}_0(\widetilde{\omega}\,R/v_{\rm F})\,\mathbb{K}_1(\widetilde{\omega}\,R/v_{\rm F})\,,\label{eq_16ne}\\
				\widetilde{\mathcal{J}}^{ \alpha\, \alpha}_{{\rm DM},y}(R)= & -2 \cos(\varphi_R)\,  \text{Im} \int_{o^+}^{\infty} \hspace*{-0.1cm} d \omega \,\widetilde{\omega} \sqrt{ \widetilde{\omega}^2-\Delta^2/4}\notag \\ {} & \hspace*{1cm} \,\times\mathbb{K}_0(\widetilde{\omega}\,R/v_{\rm F})\,\mathbb{K}_1(\widetilde{\omega}\,R/v_{\rm F})\,.\label{eq_16nf}
			\end{align}
		\end{subequations}
		Due to the symmetrical form of the states $|1\rangle$ and $|2\rangle$, the $x$ and $y$ components of the XYZ-Heisenberg interaction differ only by a sign in the off-diagonal components of the real-space Green's functions. Moreover, the DM components are shifted by the $\pi/2$ phase.
		
		For the second configuration (impurities on different sublattices), we find
		\begin{subequations}\label{eq_18n}
			\begin{align}
				\widetilde{\mathcal{J}}^{ \alpha\, \beta}_{x}(R) = {} & + \widetilde{\mathcal{J}}^{ \alpha\, \alpha}_{x}(R) \, ,\label{eq_18na}\\
				\widetilde{\mathcal{J}}^{ \alpha\, \beta}_{y}(R) = {} & - \widetilde{\mathcal{J}}^{ \alpha\, \alpha}_{y}(R) \, ,\\
				\widetilde{\mathcal{J}}^{ \alpha\, \beta}_{z}(R) = {} & - \widetilde{\mathcal{J}}^{ \alpha\, \alpha}_{z}(R) \, ,\\
				\widetilde{\mathcal{J}}^{ \alpha\, \beta}_{{\rm DM},x}(R) = {} & - \widetilde{\mathcal{J}}^{ \alpha\, \alpha}_{{\rm DM},x}(R)\, ,\\
				\widetilde{\mathcal{J}}^{ \alpha\, \beta}_{{\rm DM},z}(R) = {}  &-\widetilde{\mathcal{J}}^{ \alpha\, \alpha}_{xy}(R)\, ,\label{eq_18ne}
			\end{align}
		\end{subequations}
		stemming from the spin-orbit structure of the model, and not from the spatial symmetry of the SnTe(001) surface. 
		
		The magnetic impurities can reside on the same, Fig.~\ref{f2}(a), or different, Fig.~\ref{f2}(b), sublattices. Other configurations, such as impurities on the bonds or in the center of the unit cell, can be simply obtained by adding together the results for the RKKY interaction on the nearby sites. For example, if one magnetic impurity is located halfway between a Sn site at location $a_0\hat{y}/2$ and Te site at $-a_0\hat{y}/2$ ($a_0$ is the lattice constant), then the interaction with an impurity located at a distance $\vec{R}$ away on a Sn site is given by $\mathcal{H}^{\rm bond}_{\rm RKKY} = \mathcal{H}^{\rm SnSn}_{\rm RKKY}(\vec{R} - a_0\hat{y}/2) + \mathcal{H}^{\rm TeSn}_{\rm RKKY} (\vec{R} + a_0 \hat{y}/2)$. Note that the microscopic coupling $J$ for this case will be different from the site-located impurity, and will depend on the sublattice. Generalizations to other impurity positions are straightforward.
		\begin{figure}[t]
			\centering
			\includegraphics[width=1\linewidth]{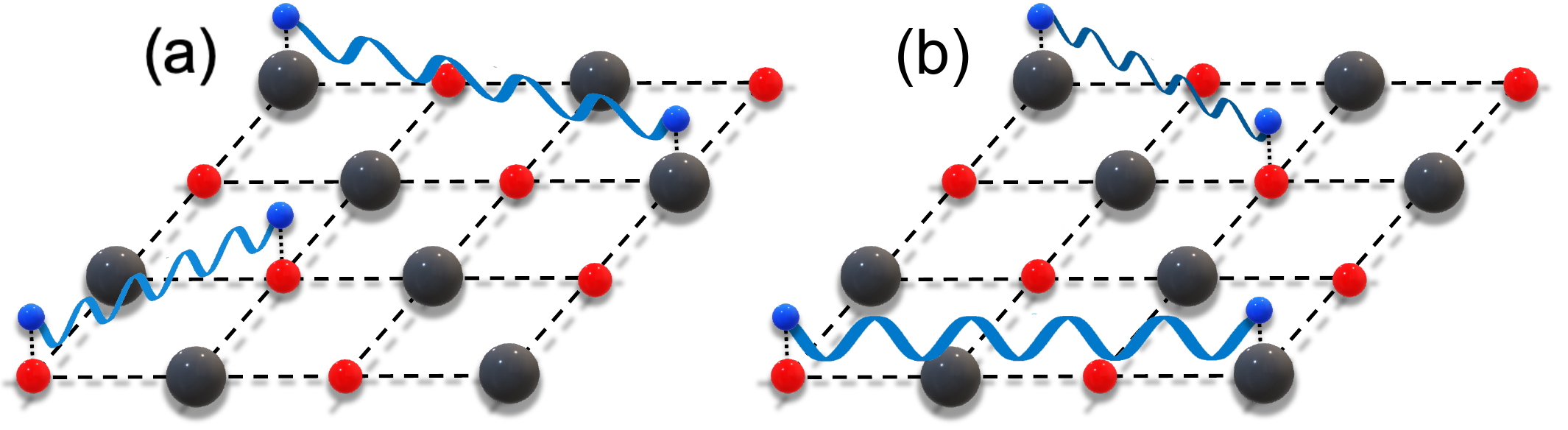}
			\caption{A simple schematic for the position of magnetic impurities on (a) the same and (b) different sublattices of the SnTe(001) surface. Cation Sn$^{2+}$, anion Te$^{2-}$, and the local magnetic impurities are, respectively, shown by the black, red, and blue spheres.} 
			\label{f2}
		\end{figure}
		
		In general, for the sublattice setup, we continue only with $\alpha_{\rm s} = +1$ in Eq.~\eqref{eq_15} henceforth for the magnetic impurities on the same TeTe sublattices. A sign change in $\widetilde{\mathcal{J}}^{\alpha \alpha}_{{\rm DM},y}$ and $\widetilde{\mathcal{J}}^{\alpha \alpha}_{xy}$ gives rise to the results for magnetic impurities on the same SnSn sublattices; for different TeSn sublattices, one can simply use Eq.~\eqref{eq_18n}. Finally, the sign of  $\widetilde{\mathcal{J}}^{\alpha \beta}_{{\rm DM},z}$ should be swapped for the results on different SnTe sublattices. 
		
		
		\section{Results and discussion}\label{s4}
		
		The role of the chemical potential $\mu$ and the optical gap $\Delta$ in various terms of the RKKY interaction form the main messages of the present work for both spintronic applications and QAHE physics. The Fermi sea can be drained out if one keeps increasing the driving strength for a fixed impurity separation $R$ and direction $\varphi_R$. This, in turn, manipulates the surface states involved in the RKKY coupling. It is very important to control the chemical potential to bring the surface states into the so-called low-energy regime. This can be tuned with, e.g., a back gate from the thin film substrates~\cite{PhysRevLett.105.176602,doi:10.1073/pnas.1713458114}.
		
		It is necessary to mention that the Floquet parameter $a/\Omega$ mainly tunes the light-matter interaction effect on the SnTe(001) surface, which should be small in the weak drive regime~(to avoid heating of the sample). It is worthwhile commenting that the alignment of magnetic moments can be influenced by both $a$ and $\Omega$. In Fig.~\ref{ap3n} of Appendix~\ref{ap_3nn}, we systematically address this matter. As soon as we turn on the light, independent of the symmetric and asymmetric contributions to the RKKY interaction, magnetic moments talk to each other only beneath the line $a \propto \sqrt{\Omega}$. The bandwidth of our model is given by $2n = 110$ meV. To have a broader range of light intensity for tuning the RKKY interaction (less zero response), we continue with an off-resonant high-frequency $\Omega = 1$ eV $\propto 270$ THz much larger than the bandwidth. Thus, the gap of the system can reach $\Delta \simeq 23$ K~\cite{doi:10.1126/science.287.5461.2237}, i.e., $a \simeq 47$ meV. However, we vary intensity $a$ up to 200 meV~(optical gap $\Delta$ up to 80 meV) in what follows.
		
		From the Eqs.~\eqref{eq_16na}-\eqref{eq_16nf} and Eqs.~\eqref{eq_18na}-\eqref{eq_18ne}, one can observe the periodicity of interactions in $\varphi_R$ -- except $\varphi_R$-independent $\widetilde{\mathcal{J}}^{ \alpha\, \alpha}_{z}(R)$ -- which provides $\mathcal{H}_{\rm RKKY}(\varphi_R) = \mathcal{H}_{\rm RKKY}(\varphi_R+\pi)$. Due to the square lattice of the host SnTe(001) surface, three angles $\varphi_R = 0$, $ \pi/4,$ and $\pi/2$ between the magnetic impurities lead to special RKKY interactions. A quick analysis results in $\widetilde{\mathcal{J}}^{ \alpha\, \alpha}_{xy}(R) = 0$ for both $0$ and $\pi/2$ and $\widetilde{\mathcal{J}}^{ \alpha\, \alpha}_{{\rm DM},x/y}(R) = 0$, respectively, for $0$ and $\pi/2$. While for $\pi/4$, we find $\widetilde{\mathcal{J}}^{\alpha\,\alpha}_{x}(R) = \widetilde{\mathcal{J}}^{\alpha\,\alpha}_{y}(R)$ and $\widetilde{\mathcal{J}}^{\alpha\,\alpha}_{{\rm DM},x}(R) = \widetilde{\mathcal{J}}^{\alpha\,\alpha}_{{\rm DM},y}(R)$. But, for noncollinear twisted alignments of magnetic moments on the surface with more adjustability, we continue with an in-plane angle, e.g., $\varphi_R = \pi/3$. We set the chemical potential and the optical gap at $\mu = 27.5$ meV and $\Delta = 45$ meV, respectively, where necessary. In the driven-doped surface, this set helps to have effective interactions because of $\mu > \Delta/2$ (see interpretation below).
		
		Figure~\ref{f4} shows the RKKY terms as a function of impurity separation $R/a_0$. For the pristine~(undriven-undoped) surface, $\mu = 0$ and $\Delta = 0$, which leads to real $\widetilde{\omega} = \omega$, the analytical solution of Eqs.~\eqref{eq_16na}-\eqref{eq_16nd} is already known~\cite{PhysRevB.102.075411,SR,PhysRevB.81.205416,PhysRevB.83.165425,PhysRevB.84.125416,Kogan} and can straightforwardly be found with the help of Eq.~\eqref{eq_d1} in Appendix~\ref{ap_3n} such that the non-zero interactions decay as $R^{-3}$, see Fig.~\ref{f4}(a). Under the same conditions, the DM terms in Eqs.~\eqref{eq_16ne} and~\eqref{eq_16nf} vanish $\widetilde{\mathcal{J}}^{\alpha \alpha}_{{\rm DM},x/y}(R) = 0$ due to the absence of imaginary energy. The physical reason for a vanishing DM interaction stems from the presence of inversion symmetry between sublattices~(absence of $\mu$) for which asymmetric alignment of spins from the interplay of diagonal and off-diagonal spin-orbital states does not take place. The presence of DM interaction due to doping has been experimentally confirmed in TI thin films~\cite{PhysRevLett.119.176809}. Interestingly, only $\widetilde{\mathcal{J}}^{\alpha \alpha}_{x}(R)$ forms an AFM order since the response from different spin-orbital states, $\{|1\rangle$ and $|2\rangle\}$ or $\{|2\rangle$ and $|1\rangle\}$, are three times larger than the same spin-orbital states, $\{|1\rangle$ and $|1\rangle\}$ or $\{|2\rangle$ and $|2\rangle\}$, resulting in $\widetilde{\mathcal{J}}^{\alpha \alpha}_{x}(R) < 0$.
		\begin{figure}[t]
			\centering
			\includegraphics[width=1\linewidth]{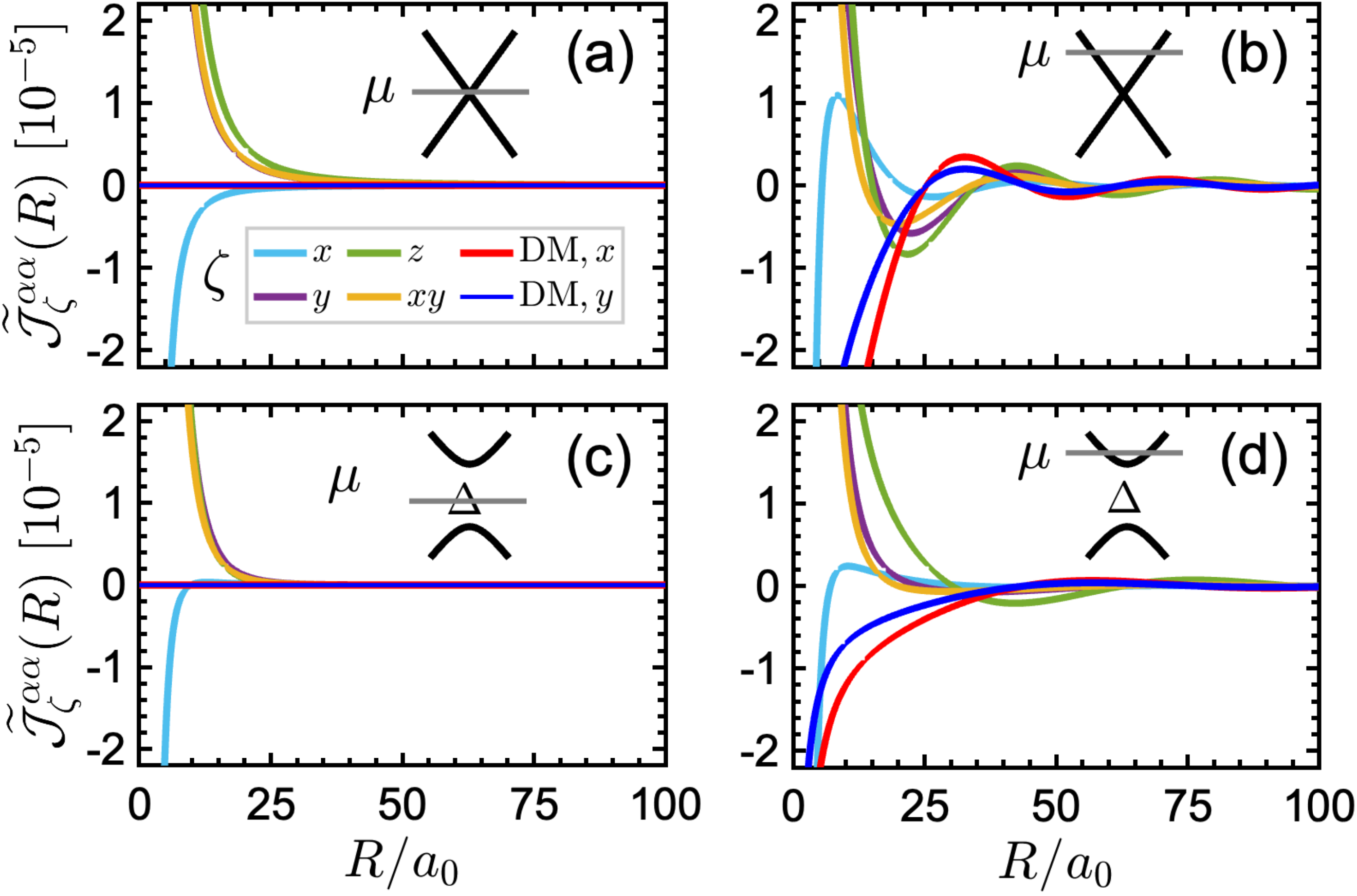}
			\caption{RKKY interactions on the SnTe(001) surface at $\varphi_R = \pi/3$ as a function of impurity separation $R$ when the surface is (a) pristine~(undriven-undoped), (b) undriven-doped, (c) driven-undoped, and (d) driven-doped. We set $\mu = 27.5$ meV and $\Delta = 45$ meV, respectively, for the chemical potential and optical gap. It can be observed that the asymmetric DM term exists only on the doped surface. Furthermore, the Floquet drive suppresses all terms due to the reduced density of mediating states.} 
			\label{f4}
		\end{figure}
		
		For the undriven-doped surface, $\mu \neq 0$ and $\Delta = 0$, which leads to imaginary $\widetilde{\omega} = \omega-{\tt i}\mu$, one needs Eq.~\eqref{eq_d2} in Appendix~\ref{ap_3n}. It is obvious that the short-range RKKY interactions are given by the power-law $R^{-3}$, while for the long-range couplings, oscillatory Meijer functions~\cite{luke1969special,10.2307/2005104} emerge, as presented in Fig.~\ref{ap3}. Our numerical data in Fig.~\ref{f4}(b) confirm this dependence. In contrast to the pristine surface, DM interactions wake up when the surface is doped for the same reason of inversion symmetry breaking between sublattices.
		
		For the driven-undoped surface, $\mu = 0$ and $\Delta \neq 0$, which leads to real $\widetilde{\omega} = \sqrt{\omega^2 +\Delta^2/4}$, with the help of Eq.~\eqref{eq_d3} in Appendix~\ref{ap_3n}, we find short-range RKKY interactions $\propto R^{-3}$, while for the long-range interactions, one needs to use $\lim_{\Delta R/v_{\rm F} \to \infty} \mathbb{K}_n(\Delta R/v_{\rm F}) \simeq \sqrt{\pi v_{\rm F}/2\Delta R} \exp(-\Delta R/v_{\rm F})$ and the Laplace method to obtain RKKY interactions $\propto (R/\Delta)^{-3/2} \exp(-2\Delta R/v_{\rm F})$~\cite{Kogan}, which are again in agreement with the numerical results in Fig.~\ref{f4}(c). For the reason mentioned before about the doping effect on the DM interaction, one still expects a vanishing DM response for the driven-undoped surface.
		
		For the driven-doped surface in Fig.~\ref{f4}(d), we consider the $\mu > \Delta/2$ regime where the chemical potential is outside the optical gap so the states added by doping play role in RKKY interactions~(for $\mu < \Delta/2$ the DM terms vanish). Although it is not easy to find analytical expressions for the driven-doped surface, we expect a modified version of Meijer functions for the long-range coupling as the interactions behave like the undriven-doped surface. However, we believe that the power-law $R^{-3}$ is still valid for short-range couplings. As both driving and doping are present, the drastic change of oscillations for all RKKY components is evident. This is a direct consequence of the optical gap for which the corresponding density of mediating states in the RKKY interaction decreases, and thus, most of the doped states are washed out. For this reason, the oscillations are accompanied by smaller amplitudes compared to the undriven-doped surface. This is mainly where our novel contribution pays off in QAHE physics since magnetic characters focus more on purely positive~(FM) and negative~(AFM) signs. 
		
		It is also worth mentioning that various FM and AFM characters emerge for symmetric XYZ-Heisenberg $\widetilde{\mathcal{J}}_{i}$ and in-plane $\widetilde{\mathcal{J}}_{xy}$ components depending on the interplay between driving and doping, as well as between the impurity separation and direction. But, for the change of signs in DM components, one needs to take into account the chirality. When DM is positive, the coupling between magnetic impurities favors clockwise rotation, meaning that it tends to align right handedly more naturally than left. Similarly, a negative DM interaction favors counter-clockwise rotation for magnetic moments.
		\begin{figure}[t]
			\centering
			\includegraphics[width=1\linewidth]{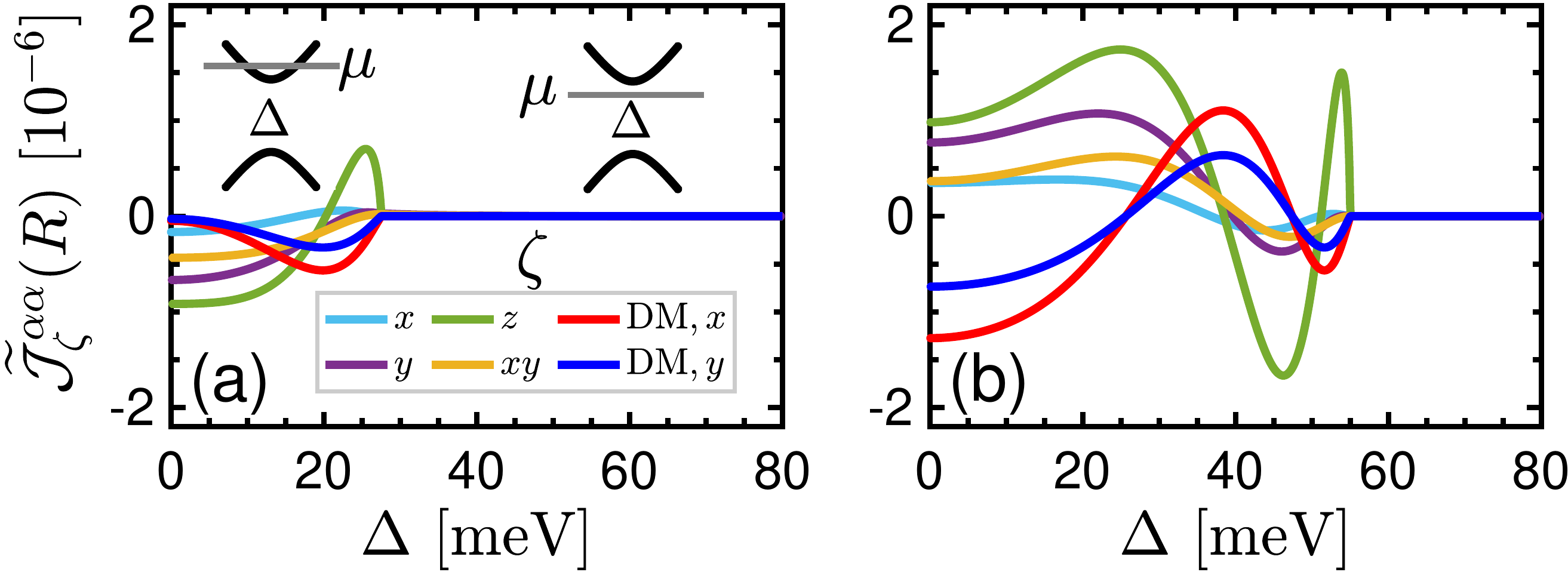}
			\caption{RKKY interactions for (a) $\mu = 13.75$ meV and (b) $\mu = 27.5$ meV at $R/a_0 = 50$ and $\varphi_R = \pi/3$ as a function of the optical gap $\Delta$ on the SnTe(001) surface. Independent of the chemical potential, RKKY responses vanish for $\Delta \geq 2 |\mu|$ once the chemical potential lies in the gap, due to their exponential decay with $R$.} 
			\label{f8}
		\end{figure}
		
		To gain further insights into the gap dependency of interactions, we proceed to plot the RKKY couplings against the optical gap $\Delta$ for $\mu = 13.75$ meV, Fig.~\ref{f8}(a), and $\mu = 27.5$ meV, Fig.~\ref{f8}(b), at $R/a_0 = 50$ and $\varphi_R = \pi/3$. As soon as we turn on the Floquet drive, all interactions start to emerge due to the presence of $\mu$. It is noteworthy that once the gap edge lies at the chemical potential level, i.e., when $\Delta = 2|\mu|$, they start to vanish because the doped states start to lie in the gap without any significant density of mediating states. If we dope the surface stronger, Fig.~\ref{f8}(b), RKKY responses become stronger compared to Fig.~\ref{f8}(a). This can be easily understood by taking into account that the density of carriers on the surface with $\mu =  27.5$ meV is larger than at $\mu = 13.75$ meV; thus, more itinerant carriers act as mediators for the indirect exchange coupling between impurities. 
		
		Next, we systematically explore the RKKY interactions of driven-doped SnTe(001) surface at fixed direction $\varphi_R = \pi/3$ and impurity separation $R/a_0 = 50$. In Fig.~\ref{f9}, at first glance, one would observe that the amplitude of oscillations increases as more surface states~(larger $\mu$) are involved in the exchange coupling. However, the gap opening reduces the rate of involved states and for a certain chemical potential $\mu$, RKKY interactions are non-zero only below $\Delta = 2 |\mu|$. For example, at $\mu = \pm \,27.5$ meV, one obtains zero RKKY responses at $\Delta \geq 50$ meV, as confirmed in Figs.~\ref{f9}(a)-(f), implying that there is no effective mediating state anymore on the surface~(at least, within our approximations) so that the RKKY responses approach zero. The red and blue colors highlight the character of FM and AFM~(clockwise and counter-clockwise spin rotations) for symmetric~(asymmetric) RKKY components, while the yellow color displays nearly zero responses.
		\begin{figure}[t]
			\centering
			\includegraphics[width=1\linewidth]{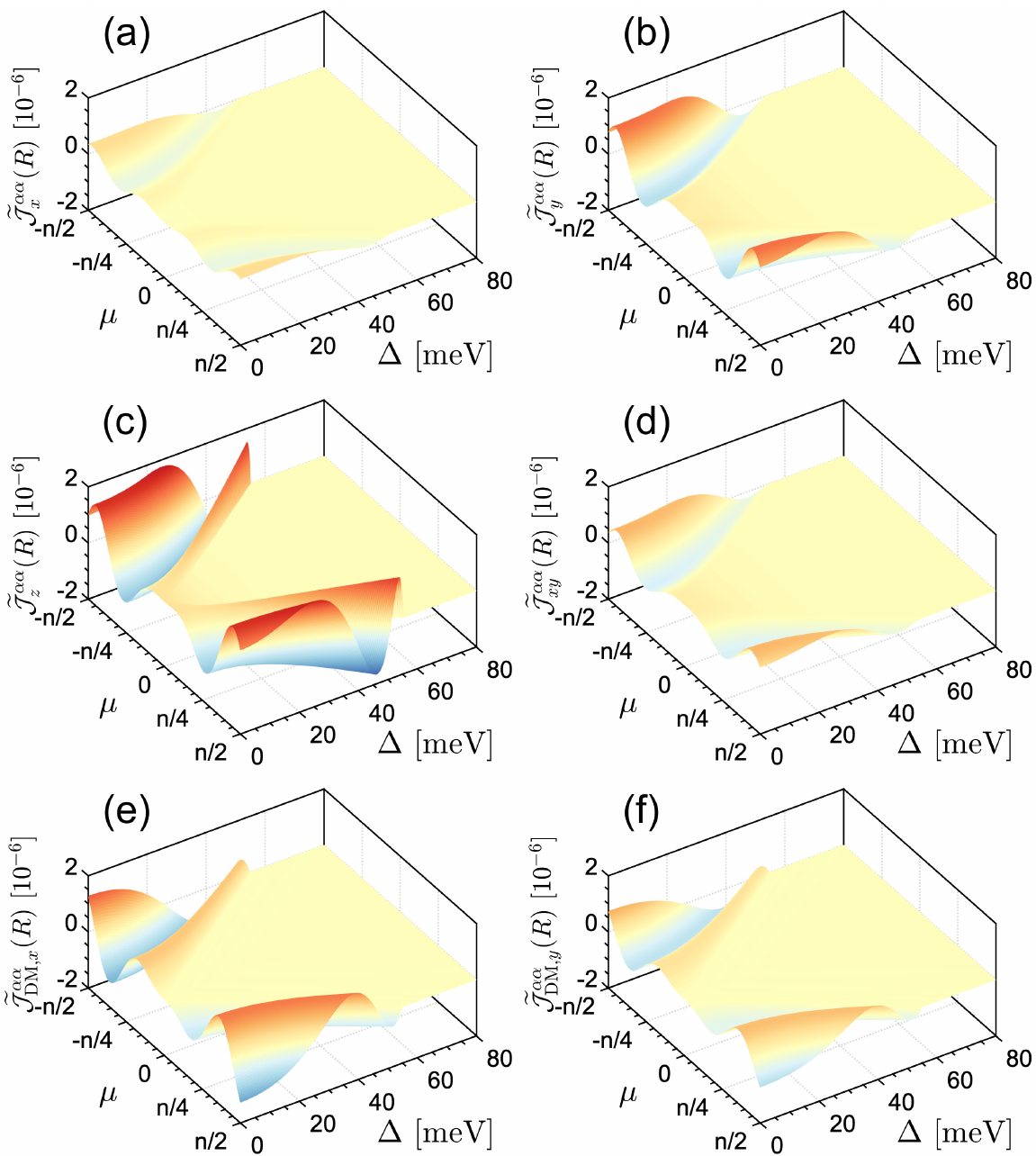}
			\caption{RKKY interactions at $\varphi_R = \pi/3$ and $R/a_0 = 50$ as a function of both optical gap $\Delta$ and chemical potential $\mu$ on the driven-doped SnTe(001) surface. Red~(blue) color refers to FM~(AFM) order in (a)-(d), while they refer to clockwise~(contour-clockwise) spin rotation in the DM terms of (e) and (f). The yellow color depicts nearly zero RKKY response. The dominant contribution to the RKKY interaction belongs to the $z$~($x$)-component of symmetric~(asymmetric) interaction, i.e., $\widetilde{\mathcal{J}}^{\alpha \alpha}_z$~($\widetilde{\mathcal{J}}^{\alpha \alpha}_{{\rm DM},x}$). As expected, no response appears for $\Delta \geq 2 |\mu|$ due to the vanishing density of mediating states.} 
			\label{f9}
		\end{figure}
		
		Additionally, the dominant oscillation amplitude belongs to the $z$-component of the XYZ-Heisenberg, though the $x$-component of the DM term is the next prevalent among others. As mentioned before, DM interaction enables the system to be applicable for emergent spintronic applications~\cite{PhysRevLett.119.176809,Wakatsuki2015,Fert2013,Sampaio2013}. This does not occur on the surface of conventional TI thin films~\cite{PhysRevB.96.024413}. Therefore, our large DM compatible with $\widetilde{\mathcal{J}}^{\alpha \alpha}_z(R)$ is a very useful feature of TCIs.
		
		It is also worth exploring the intrinsic electron-hole symmetry on the driven-doped SnTe(001) surface. The electron-hole symmetry is valid for all symmetric interactions, XYZ-Heisenberg and in-plane $xy$, independent of doping and optical gap~(also independent of the impurity separation and position of impurities -- not shown here). For the DM interaction at a given direction, one immediately extracts the following relation
		\begin{equation}\label{eq_18}
			\widetilde{\mathcal{J}}^{\alpha \alpha}_{{\rm DM},x/y}(-\mu, \Delta) =  -\widetilde{\mathcal{J}}^{\alpha \alpha}_{{\rm DM},x/y}(+\mu, \Delta) \, ,
		\end{equation}
		which has already been confirmed experimentally on the surface of TI thin films~\cite{PhysRevLett.119.176809}, as also shown in Figs.~\ref{f9}(e) and~\ref{f9}(f). It has an opposite sign when the chemical potential cuts through the upper or lower band. We emphasize that the presence of inversion symmetry between sublattices and the corresponding spin-orbital states is the main origin of the above relationship.
		
		For the experimental perspectives, one would extract the following critical Floquet driving amplitude $E^{\rm c}_0$ (considering an arbitrary driving frequency $\Omega$) in the low-energy regime of SnTe(001) surface above which a nearly zero RKKY coupling is achieved:
		\begin{equation}\label{eq_30}
			E^{\rm c}_0 = {} \frac{\sqrt{n}}{e v_{\rm F}}\Omega^{3/2}\, .
		\end{equation}
		For instance, in our case, for the selected light frequency of $\Omega \simeq 1$ eV, the Fermi velocity of $v_{\rm F} \simeq 2$ eV. \AA, and $n = 55$ meV, one finds $E^{\rm c}_0 \simeq 10^9$ V/m, which is nowadays easily realizable in experiment~\cite{doi:10.1126/science.1239834,PhysRevResearch.2.043408,McIver2020}. Having this information, we would suggest using a light intensity below $E^{\rm c}_0$ in experiments so tuning the RKKY coupling with light can play role in determining the prerequisites for spintronic applications and QAHE. Experimentally, single-atomic magnetometry and magnetotransport through scanning tunneling microscopy and angle-resolved photoemission spectroscopy can measure the RKKY interaction~\cite{PhysRevLett.91.116601,Khajetoorians2012,Zhou2010}. 
		
		
		\section{Outlook}\label{s5n}
		
		Before closing the discussion, let us briefly discuss possible avenues for upcoming research. The present study for the ``weak'' light-induced RKKY interaction, which still reports applicable findings, can also be addressed for the ``strong'' driving regime. First, in the strong driving regime, the full Floquet Hamiltonian is required for further photon processes~\cite{PhysRevB.85.155449}; second, the nonlinear corrections should be solved numerically~\cite{PhysRevB.85.155449,PhysRevResearch.2.033228}. The extended continuous strong driving must inevitably lead to heating, which without remediation would destroy the sample. We can propose a heat sink, e.g., a high-quality Al~\cite{yarmohammadi2020dynamical}, which maintains a low system temperature despite the Floquet drive. Beyond the weak driving regime, our calculations could also be done for a ``linearly'' polarized light if one considers the $n$th-order Bessel functions for the corrections up to infinite orders~\cite{PhysRevA.81.022117}. 
		
		To address the low-frequency limit beyond our high-frequency expansion, band crossings should also be considered in the theory; this requires another expansion of the effective Floquet Hamiltonian and extra terms will be generated. Additionally, the finite temperature effects can also be explored for both short- and long-range couplings within the finite-temperature self-consistent field approximation~\cite{PhysRevB.98.064425,PhysRevB.95.155414}. 
		
		Note that the drive used to tune the RKKY interaction also opens up a topological gap in the surface Dirac cones. With four Dirac cones of identical chirality, it will correspond to Chern number 2, suggesting that TCIs may also be a useful playground for light-induced anomalous quantum Hall physics. Unlike existing two-dimensional systems, such as graphene~\cite{McIver2020}, we propose to realize the gapped TCI surface states by continuous, moderate amplitude optical driving, which allowed the realization of an effective Fermi surface earlier in the paper, Sec~\ref{s4}. This also implies the quasi-static topology of the surface electronic states, which will be an interesting topic for future exploration. However, unlike graphene, this topological state is complicated by the presence of gapless edge states on the other [non-(001)] surfaces as well as the bottom (001) surface of the material. We leave an analysis of topological transport in this geometry for future work.
		
		
		\section{Conclusions}\label{s6}
		
		Exploring the physics of magnetically doped topological systems in tuning QAHE as well as for the spintronics community has triggered interest in condensed matter physics. Recently, experimental photonic platforms have also become increasingly urgent for tuning the properties of materials. To contribute to these, we have employed the isotropic optically driven continuum model for Dirac cones with nontrivial spin textures on the doped SnTe(001) surface as a well-known TCI to investigate the (quasi)out-of-equilibrium physics of the RKKY interaction between two magnetic impurities. This, in turn, aims at providing spintronic applications as well as making the TCI-based QAHE strong. To tune the features, in particular, we have focused on the weak driving effects and off-resonant regime between the light and bandwidth of the Dirac cones using the van Vleck inverse-frequency expansion. We make use of the bare spin-orbital states of each sublattice, resulting in noncollinear twisted RKKY interaction comprising of XYZ-Heisenberg, symmetric in-plane, and asymmetric DM terms. 
		
		Preliminary analyses highlighted the $z$~($x$)-component of the XYZ-Heisenberg~(DM) interaction as the first~(second) dominant contribution to the total RKKY interaction. Moreover, depending on the position of magnetic impurities, chemical potential, and the optical gap, the driven-doped SnTe(001) surface reaches various modulations of FM and AFM characters for the symmetric RKKY terms as well as of the clockwise and counter-clockwise spin rotations for the asymmetric DM ones. A systematic analysis of the RKKY coupling on the interplay between doping and driving demonstrated that the efficient range of chemical potential in controlling the amplitudes belongs to the strengths outside the optical gap. The optical gap~(Floquet drive) in the presence of doping leads to a reduction of the RKKY amplitudes because of the decreased total density of mediating states. This is where we propose a Floquet drive to make the TCI-based QAHE strong when the surface is doped.  Alongside, we find non-zero DM interactions only for the doped surface because of inversion symmetry breaking between sublattices. Moreover, to evaluate the intrinsic electron-hole symmetry of the system, we found that the symmetry is broken because of the DM term. 
		
		Providing reasonable light intensity and frequency compatible with the low-energy Dirac bands of the SnTe(001) surface, insights from the present work are discussed for feasible experimental observations. Finally, we would mention that the large DM term on the surface of TCIs is highly desirable for spintronic applications, highlighting their additional usefulness compared to TIs due to the presence of multiple Dirac cones. 
		
		
		\section*{Acknowledgments}
		
		M.Y. greatly thanks Wang-Kong Tse for helpful discussions. This work was supported by the National Science Foundation through award numbers DMR-1945529, PHY-1607611, NSF PHY1748958, and the Welch Foundation through award number AT-2036-20200401. This project was funded by The University of Texas at Dallas Office of Research and Innovation through the SPIRe program. M.B.~was supported by the Marie Sk\l{}odowska-Curie grant agreement No 890711 (until 01.09.2022).
		
		
		\appendix
		\renewcommand\thefigure{\thesection.\arabic{figure}}

		\section{Components of the real-space Green's function}\label{ap_1}
		
		The real-space Green's functions of hybrid states in terms of bare sublattices considering corresponding spins requires the following definition $G^{\sigma \sigma'}_{\alpha \beta}(\mathcal{E},\vec{R}) = \sum_{\gamma,\eta} u^*_{\alpha,\gamma,\sigma}u_{\beta,\gamma,\sigma'} \mathcal{G}_{\gamma \eta}(\mathcal{E},\vec{R})$, where components are given by [in the following we omit $(\dots)$, so $\mathcal{G}_{\gamma \eta}$ means $\mathcal{G}_{\gamma \eta}(\mathcal{E}, \vec{R})$]
		\begin{subequations}\label{eq1_ap_1}
			\begin{align}
				G^{\uparrow \uparrow}_{\rm Te\, Te} = {} & \, \mathcal{G}_{11}/2\,,\quad 
				G^{\uparrow \downarrow}_{\rm Te\, Te} = {} \, \mathcal{G}_{12}/2\,,\quad  \\
				G^{\downarrow \uparrow}_{\rm Te\, Te} = {}  &\, \mathcal{G}_{21}/2\,,\quad 
				G^{\downarrow \downarrow}_{\rm Te\, Te} = {} \, \mathcal{G}_{22}/2\,,\\
				G^{\uparrow \uparrow}_{\rm Te\, Sn} = {} & \, \mathcal{G}_{12}/2\,,\quad 
				G^{\uparrow \downarrow}_{\rm Te\, Sn} = {}  \, \mathcal{G}_{11}/2\,,\quad \\
				G^{\downarrow \uparrow}_{\rm Te\, Sn} = {}& \, \mathcal{G}_{22}/2\,,\quad 
				G^{\downarrow \downarrow}_{\rm Te\, Sn} = {}  \, \mathcal{G}_{21}/2\,,\\
				G^{\uparrow \uparrow }_{\rm Sn\, Te}= {} & \, \mathcal{G}_{21}/2\,,\quad 
				G^{\uparrow \downarrow}_{\rm Sn\, Te} = {}  \, \mathcal{G}_{22}/2\,,\quad  \\
				G^{\downarrow \uparrow}_{\rm Sn\, Te} = {}& \, \mathcal{G}_{11}/2\,,\quad 
				G^{\downarrow \downarrow}_{\rm Sn\, Te} = {}  \, \mathcal{G}_{12}/2\, ,\\
				G^{\uparrow \uparrow}_{\rm Sn\, Sn} = {} & \, \mathcal{G}_{22}/2\,,\quad 
				G^{\uparrow \downarrow}_{\rm Sn\, Sn} = {}  \, \mathcal{G}_{21}/2\,,\quad \\
				G^{\downarrow \uparrow}_{\rm Sn\, Sn} = {}  &\, \mathcal{G}_{12}/2\,,\quad
				G^{\downarrow \downarrow }_{\rm Sn\, Sn}= {}  \, \mathcal{G}_{11}/2\, .
			\end{align}
		\end{subequations}
		
		\section{Components of the spin susceptibility}\label{ap_2}
		
		The spin susceptibility is described by $\chi^{\alpha \beta}_{lj}(\,\vec{R}\,) = -\frac{2}{\pi}\, \Im \int^{\mu}_{-\infty} d\mathcal{E}\, \mathbb{F}^{\alpha \beta}_{lj}(\mathcal{E},\vec{R})$ with the following expressions for the first configuration~(impurities on the same sublattices)
		\begin{subequations}\label{eq1_ap_2}
			\begin{align}
				\mathbb{F}^{\rm \alpha\,\alpha}_{xx} = {} & \frac{|f(\vec R)|^2}{4} \left[2\,\mathcal{G}_{11} \mathcal{G}_{22} - \mathcal{G}^2_{12}- \mathcal{G}^2_{21}\right] \, , \\
				\mathbb{F}^{\rm \alpha\,\alpha}_{yy} =  {} & \frac{|f(\vec R)|^2}{4} \left[2\,\mathcal{G}_{11} \mathcal{G}_{22}+ \mathcal{G}^2_{12} +\mathcal{G}^2_{21}\right] \, ,\\
				\mathbb{F}^{\rm \alpha\,\alpha}_{zz} =  {}  &\frac{|f(\vec R)|^2}{4} \left[2\,\mathcal{G}_{12} \mathcal{G}_{21}+ \mathcal{G}^2_{11}+ \mathcal{G}^2_{22}\right] \,, \\
				\mathbb{F}^{\rm \alpha\,\alpha, \alpha_{\rm s}}_{xy/yx} =  {}  &{\tt i}\alpha_{\rm s}\,\frac{|f(\vec R)|^2}{4} \left[\mathcal{G}^2_{21}-\mathcal{G}^2_{12}\right]  \, ,\\
				\mathbb{F}^{\rm \alpha\,\alpha,\alpha_{\rm s}}_{xz/zx} =  {} & \pm\alpha_{\rm s}\,\frac{|f(\vec R)|^2}{4} \left[\mathcal{G}_{11}+\mathcal{G}_{22}\right] \left[ \mathcal{G}_{21}-\mathcal{G}_{12}\right] \, , \\
				\mathbb{F}^{\rm \alpha\,\alpha}_{yz/zy} =  {}  &\mp {\tt i}\frac{|f(\vec R)|^2}{4} \left[\mathcal{G}_{11}+\mathcal{G}_{22}\right] \left[ \mathcal{G}_{12}+ \mathcal{G}_{21}\right] \, .
			\end{align}
		\end{subequations}
		For the second configuration~(impurities on different sublattices), we acheive
		\begin{subequations}\label{eq1_ap_3}
			\begin{align}
				\mathbb{F}^{ \alpha\,\beta/\beta\,\alpha}_{xx} =  {} & \frac{|f(\vec R)|^2}{4} \left[2\,\mathcal{G}_{11}\mathcal{G}_{22}-\mathcal{G}^2_{12} -\mathcal{G}^2_{21}\right]   \, , \\
				\mathbb{F}^{ \alpha\,\beta/\beta\,\alpha}_{yy} =  {}  &\frac{|f(\vec R)|^2}{4} \left[-2\,\mathcal{G}_{11}\mathcal{G}_{22}-\mathcal{G}^2_{12} -\mathcal{G}^2_{21}\right] \, ,\\
				\mathbb{F}^{ \alpha\,\beta/\beta\,\alpha}_{zz} =  {}  &\frac{|f(\vec R)|^2}{4} \left[-2\,\mathcal{G}_{12}\mathcal{G}_{21}-\mathcal{G}^2_{11} -\mathcal{G}^2_{22}\right] \, ,\\
				\mathbb{F}^{\rm \alpha\,\beta,\alpha_{\rm d}}_{xy/yx} =  {} & \pm {\tt i} \alpha_{\rm d}\, \frac{|f(\vec R)|^2}{4} \left[\mathcal{G}^2_{12}-\mathcal{G}^2_{21}\right]\, ,  \\
				\mathbb{F}^{\rm \alpha\,\beta,\alpha_{\rm d}}_{xz/zx} =  {}  & \alpha_{\rm d}\,  \frac{|f(\vec R)|^2}{4} \left[\mathcal{G}_{11}+\mathcal{G}_{22}\right]\left[\mathcal{G}_{12}-\mathcal{G}_{21}\right]  \, , \\
				\mathbb{F}^{\rm \alpha\,\beta}_{yz/zy} =  {} & \pm {\tt i} \frac{|f(\vec R)|^2}{4} \left[\mathcal{G}_{11}+\mathcal{G}_{22}\right]\left[\mathcal{G}_{12}+\mathcal{G}_{21}\right] \,.
			\end{align}
		\end{subequations}
		
			\section{Beating oscillation of RKKY interaction}\label{ap_1n}
		
		\setcounter{figure}{0} 
		\begin{figure}[b]
			\centering
			\includegraphics[width=0.9\linewidth]{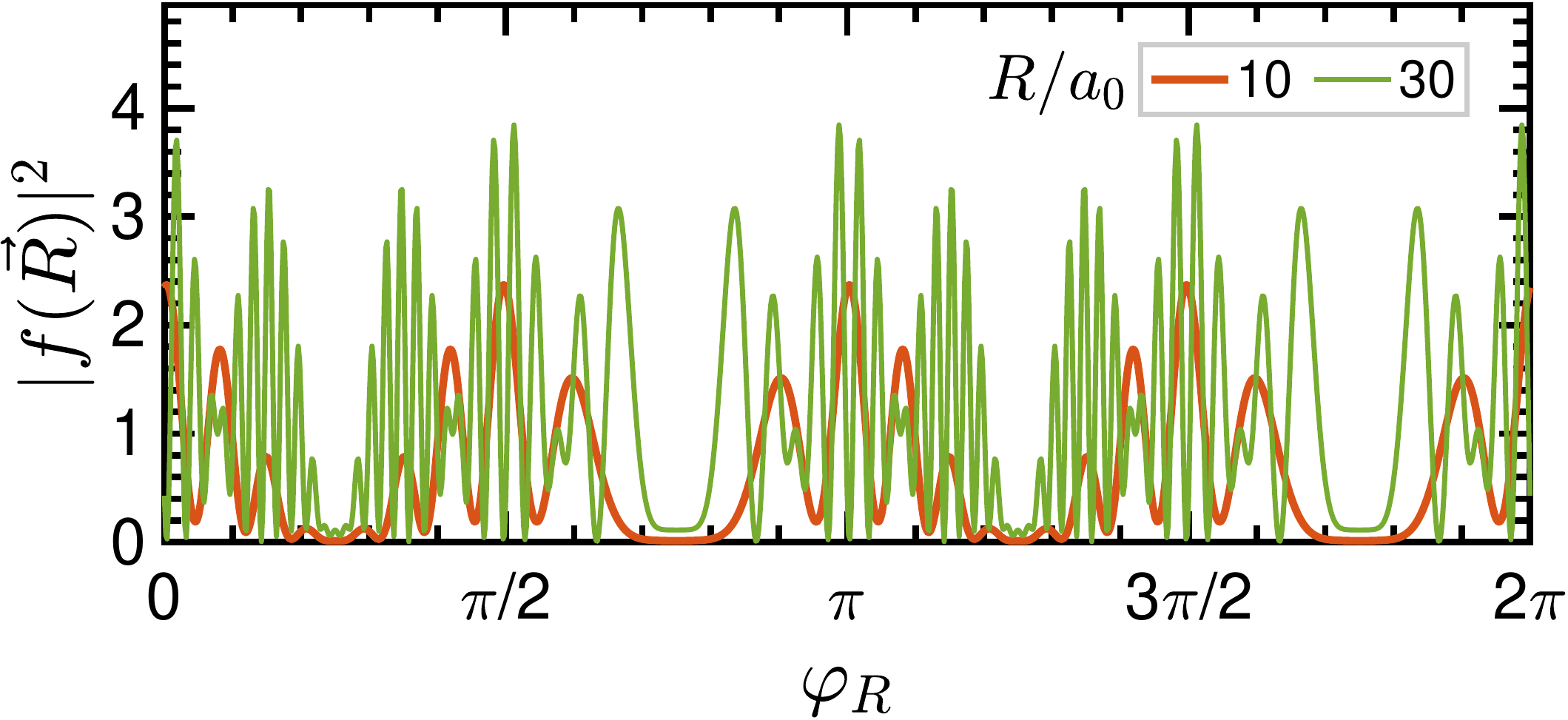}
			\caption{Beating type of RKKY oscillations due to the multiple Dirac cones on the SnTe(001) surface, Eq.~\eqref{eq_15n}, for impurity separations $R/a_0 = 10$ and 30 as a function of the polar angle $\varphi_R$.} 
			\label{fc1}
		\end{figure}
		Here, we aim at showing the beating type of oscillations from multiple Dirac cones on the SnTe(001) surface, see Eq.~\ref{eq_15n} for $|f(\vec{R})|^2$, as shown in Fig.~\ref{fc1}. The momenta contributed to this beating function are $X_{1/2} \pm \Lambda_{x/y}$ and $2\Lambda_{x/y}$ on the SBZ~\cite{PhysRevB.102.075411}.
		
		\section{Floquet engineering of magnetic moments alignment on the $\rm SnTe$(001) surface}\label{ap_3nn}
		
		\setcounter{figure}{0} 
		\begin{figure}[t]
			\centering
			\includegraphics[width=0.9\linewidth]{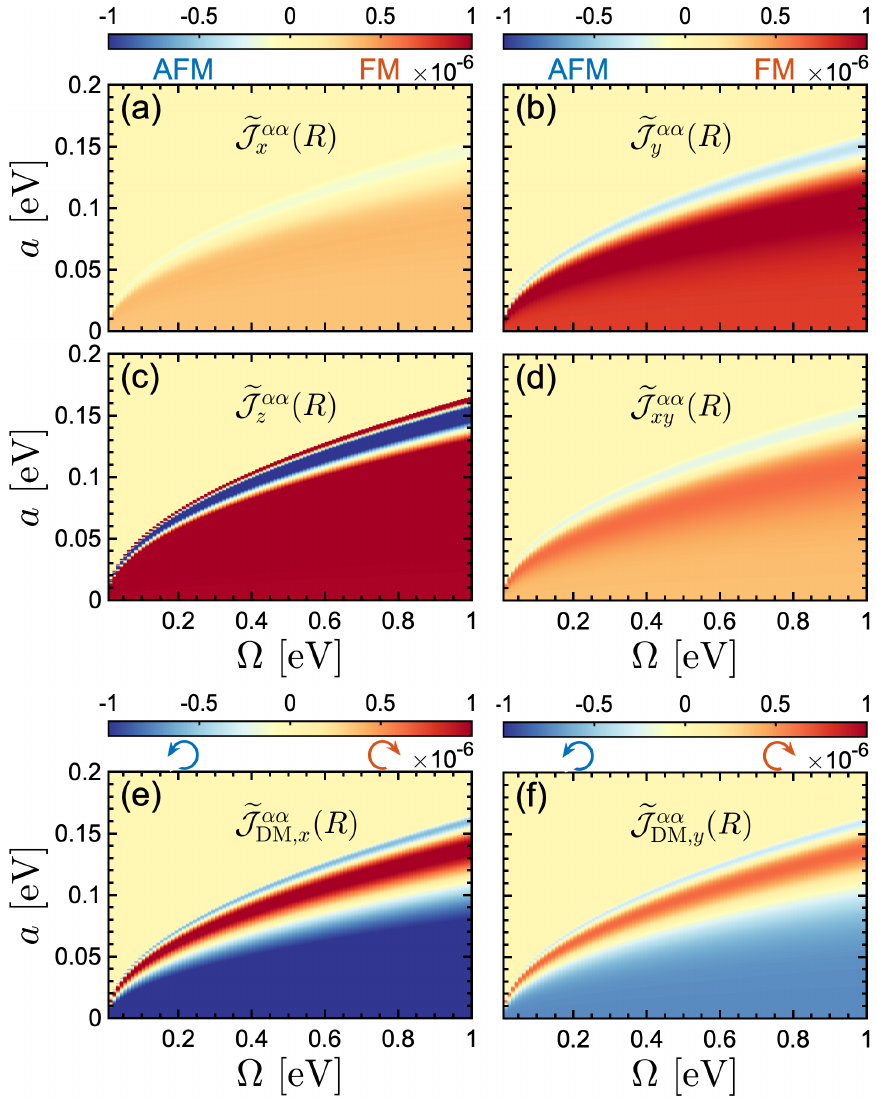}
			\caption{Magnetic phase diagram of RKKY interactions on the driven-doped SnTe(001) surface at $R/a_0 = 50$ and $\varphi_R = \pi/3$. Depending on the interplay between light intensity $a$ and frequency $\Omega$, various FM/AFM and clockwise/contour-clockwise characters can be seen for symmetric and asymmetric components, respectively. Also, $\widetilde{\mathcal{J}}^{\alpha \alpha}_z(R) > \widetilde{\mathcal{J}}^{\alpha \alpha}_{{\rm DM},x}(R)$ holds true.} 
			\label{ap3n}
		\end{figure}
		An important motivation for Fig.~\ref{ap3n} presented here originates from the spintronics community where the quest for the control of magnetic phases is still active. Among the fundamental open questions is whether the magnetic moments alignment on the SnTe(001) surface with a nontrivial spin texture is adjustable through the interplay between the intensity and frequency of light. This is somehow important for the Floqeut engineering of surface. As can be seen, phase diagrams for symmetric and asymmetric contributions to the RKKY interaction highlight the characters of FM/AFM~(for symmetric contributions in Figs.~\ref{ap3n}(a)-(d)) as well as clockwise/contour-clockwise~(for asymmetric contributions in Figs.~\ref{ap3n}(e)-(f)) in different controlled ways, characterized by red and blue colors. In a certain regime above the line $a \propto \sqrt{\Omega}$, the RKKY interaction vanishes. Getting away from the alignments, we have the relation $\widetilde{\mathcal{J}}^{\alpha \alpha}_z(R) > \widetilde{\mathcal{J}}^{\alpha \alpha}_{{\rm DM},x}(R)$ between the strongest symmetric and asymmetric RKKY components.
		
		\section{Solution of integrals in Eq.~\eqref{eq_16} for undriven-undoped, undriven-doped, and driven-undoped $\rm SnTe$(001) surface}\label{ap_3n}
		
		\setcounter{figure}{0} 
		\begin{figure}[b]
			\centering
			\includegraphics[width=0.9\linewidth]{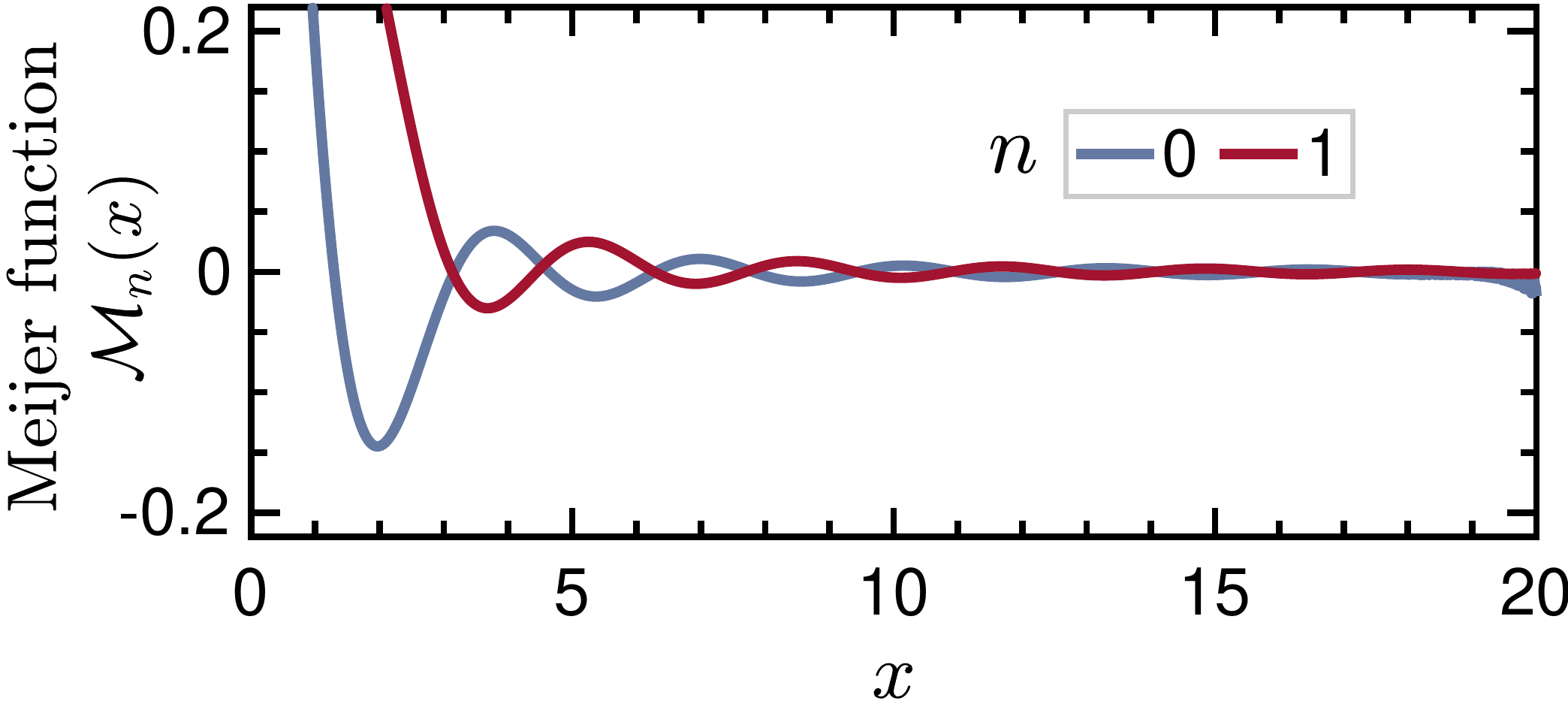}
			\caption{Oscillatory Meijer functions, Eq.~\eqref{eq_d2}, versus $x = \mu R/v_{\rm F}$ for undriven-doped SnTe(001) surface, resulting in oscillatory RKKY interactions in Fig.~\ref{f4}(b).} 
			\label{ap3}
		\end{figure}
		In this part, we present the analytical expressions for the integrals in Eq.~\eqref{eq_16}. For the pristine surface, we use the following relation for $n = \{0,1\}$:
		\begin{equation}\label{eq_d1}
			{\rm Im} \int^{\infty}_{o^+} {\tt i} d \omega\, \omega^2 \mathbb{K}^2_n(\omega R/v_{\rm F}) = + (2n+1)\frac{\pi^2 v^3_{\rm F}}{32 R^3}\, .
		\end{equation}
		For the undriven-doped surface, the following relations are required to understand the short- and long-range responses:
		\begin{subequations}\label{eq_d2}
			\begin{align}
				&\mathbb{K}_n({\tt i}u) = {}  (-1)^{n+1}\frac{\pi}{2} e^{{\tt i}n\pi/2} \big[\mathbb{Y}_n(u) + {\tt i}\mathbb{J}_n(u) \big]\, ,\\
				&	{\rm Im} \int^{\infty}_{o^+} {\tt i} d \omega [\omega-{\tt i}\mu]^2 \mathbb{K}^2_n([\omega-{\tt i}\mu] R/v_{\rm F}) =	{}\notag \\{} & + (2n+1)\frac{\pi^2 v^3_{\rm F}}{32 R^3} - (-1)^n \frac{\pi^2\mu^3}{4 \sqrt{\pi}} \underbrace{\MeijerG*{2}{1}{2}{4}{-\tfrac{1}{2}, \tfrac{1}{2}}{0,n,-\tfrac{3}{2},-n}{\frac{\mu^2R^2}{v^2_{\rm F}}}}_{\mathcal{M}_n(\mu R/v_{\rm F})},
			\end{align}
		\end{subequations}where $\mathcal{M}_n(x)$ is the special Meijer function~\cite{luke1969special,10.2307/2005104}, as shown in Fig.~\ref{ap3}, which leads to oscillatory RKKY interactions in Fig.~\ref{f4}(b).
		
		For the driven-undoped surface, we use the following mathematical identity~\cite{prudnikov1986integrals}
		\begin{equation}\label{eq_d3}
			\begin{aligned}
				&\int_{-\infty}^{0} x^{j-1} d x \,\mathbb{K}_n(a x) \mathbb{K}_m(a x) = {}  \frac{2^{j-3}}{a^j \Gamma(j)} \Gamma([j+n+m]/2)\\ {} &
				\quad \Gamma([j+n-m]/2) \Gamma([j-n+m]/2) \Gamma([j-n-m]/2)\, ,
			\end{aligned}
		\end{equation}where $\Gamma(\dots)$ is the special Gamma function.
	}
	
	\twocolumngrid
	\bibliography{bib.bib}
\end{document}